\documentclass{iopart}
\usepackage{iopams} 
\usepackage{epsfig}
\usepackage{amssymb}
\usepackage[T1]{fontenc}
\usepackage[utf8]{inputenc}
\usepackage{ae,aecompl}
\usepackage{lmodern}

\usepackage{color}
\definecolor{darkblue}{rgb}{0,0,0.6}
\definecolor{darkred}{rgb}{0.6,0,0}
\definecolor{darkgreen}{rgb}{0,0.6,0}
\definecolor{darkyellow}{rgb}{0.5,0.5,0}
\definecolor{darkpurple}{rgb}{0.6,0,0.6}
\definecolor{darkgrey}{rgb}{0.4,0.4,0.4}
\usepackage{graphicx}
\usepackage[colorlinks=true,urlcolor=darkblue,citecolor=darkblue,linkcolor=darkred,hyperfootnotes=false]{hyperref}
\usepackage{url}

\newcommand{\dd}{\textnormal{d}}

\newcommand{\ee}{{\rm e}}
\newcommand{\eee}{{\rm e}}
\newcommand{\sym}{{\rm sym}}

\newcommand{\evol}{W}
\newcommand{\eqref}{\eref}
\newcommand{\text}{\textnormal}
\newcommand{\tomax}{0}
\newcommand{\rc}{{\rm c}}

\usepackage[title,titletoc]{appendix}

\usepackage{longtable}

\begin{document}

\title{Finite-size effects in a mean-field kinetically constrained model: dynamical glassiness and quantum criticality}

\author{Takahiro Nemoto$^1$\footnote{\mailto{nemoto@ton.scphys.kyoto-u.ac.jp}}, Vivien Lecomte$^2$\footnote{\mailto{vivien.lecomte@univ-paris-diderot.fr}}, Shin-ichi Sasa$^1$, and Fr\'ed\'eric van Wijland$^3$}

\address{$^{1}$Division of Physics and Astronomy, Graduate School of Science, Kyoto University, Kitashirakawa-Oiwakecho, Sakyo-ku, Kyoto 606-8502, Japan.\\
$^{2}$Laboratoire de Probabilit\'es et Mod\`eles Al\'eatoires, UMR 7599 CNRS/P7, Universit\'e Paris Diderot, 5 rue Thomas Mann, 75205 Paris cedex 13, France.\\
$^{3}$Laboratoire Mati\`ere et Syst\`emes Complexes, UMR 7057 CNRS/P7, Universit\'e Paris Diderot, 5 rue Thomas Mann, 75205 Paris cedex 13, France}

\begin{abstract}On the example of a mean-field Fredrickson-Andersen kinetically constrained model, we focus on the known property that equilibrium dynamics take place at a first-order dynamical phase transition point in the space of time-realizations. We investigate the finite-size properties of this first order transition. By discussing and exploiting a mapping of the classical dynamical transition --~an argued glassiness signature~-- to a first-order quantum transition, we show that the quantum analogy can be exploited to extract finite-size properties, which in many respects are similar to those in genuine mean-field quantum systems with a first-order transition. We fully characterize the finite-size properties of the order parameter across the first order transition.
\end{abstract}

\setcounter{tocdepth}{2}
\tableofcontents

\maketitle

\section{Introduction}
 \bigskip
In the realm of the modeling of glassy materials, the idea that metastable states are responsible for the slowing down of the dynamics is an accepted statement. What is however still a matter of debate is whether metastability arises from the wells and valleys of some underlying complex energy landscape, or whether it is dynamically induced by the evolution within the latter energy landscape. Our purpose in this work is not to fuel this debate, but to investigate some refined features of dynamics induced metastability. A family of model systems exhibiting such glassy-like properties is that of kinetically constrained models (KCMs) as presented in the review by Ritort and Sollich~\cite{doi:10.1080/0001873031000093582}, or more recently by Garrahan, Sollich and Toninelli~\cite{garrahan_kinetically_2010}. When adopting the standpoint of Ruelle, Sinai and Bowen's thermodynamic formalism~\cite{ruelle_thermodynamic_2004,sinai_gibbs_1972,bowen_equilibrium_2008}, a Gibbs ensemble construction based on time 
realizations rather than on instantaneous configurations, it can be seen that the equilibrium 
dynamics of KCMs take place at a first-order critical point. In a nutshell, trajectories over a large time interval are ordered according to a prescribed value of some meaningful space and time extensive physical observable. The general theory within the framework of Markov dynamics was described by Lecomte {\it et al.}~\cite{PhysRevLett.95.010601}, and the idea that this could be relevant to glassy systems was initially put forward by Merolle {\it et al.}~\cite{merolle_spacetime_2005}. For KCMs, which are lattice systems with discrete degrees of freedom whose evolution rules are encoded in a master equation, the dynamic evolution rules satisfy the detailed balance with respect to an equilibrium distribution for independent degrees of freedom, but the dynamics itself is highly correlated. In KCMs the degrees of freedom are represented by Ising spins or by local occupations numbers (0 or 1) which physically represent local coarse-grained patches of activity. A simple physical observable that we will use to 
measure the 
overall activity of a time-realization is the number of configuration changes that have taken place over the observation interval. Using the latter quantity to characterize time realizations has revealed that, in KCMs, equilibrium trajectories can be divided into two coexisting groups, those which display a finite activity, and those which are basically frozen in. Within the thermodynamic formalism, this phenomenon is called a dynamical first-order transition, and it can be cast within the same mathematics as that used for ordinary equilibrium (liquid-gas) first-order transition. A full account can be found in Garrahan {\it et al.}~\cite{garrahanjacklecomtepitardvanduijvendijkvanwijland,garrahan_first-order_2009}. In existing simulations on KCMs that probe this first-order transition scenario, one key difficulty is to overcome the critical point the vicinity of which is responsible for huge numerical difficulties (slowing down, large fluctuations, insufficient numerical sampling). These finite-size effects 
turn out to be even more serious in realistic 
atomistic models of glasses~\cite{hedges_dynamic_2009, pitard_dynamic_2011}. Simulations in these systems are necessarily carried in both finite time and finite size. However, while first order dynamical transitions can be found in systems with few degrees of freedom (such as the simple harmonic oscillator of Whitelam and Garrahan~\cite{doi:10.1021/jp037653x} using the time integrated potential energy in lieu of the activity to probe time realizations), those found in KCMs only emerge in the large system size limit. It is therefore crucial to master the behavior of infinite-time but finite-size effects to properly analyze numerical data.

In an effort to address some of these issues, Bodineau and Toninelli~\cite{bodineautoninelli} proved the existence of a surface tension in the one-dimensional East and Fredrickson-Andersen (FA) models at the coexistence point. In the same vein, an effective interface model was then used by Bodineau, Lecomte and Toninelli~\cite{bodineaulecomtetoninelli} to describe finite-size effects in the one-dimensional FA model. A fully solvable model for dynamical first-order phase transitions still hasn't been put forward (unlike second-order phase transitions~\cite{1751-8121-45-17-175001}). It is the purpose of this work to show that, at the price of losing finite-dimensional effects (such as surface tension), the fully-connected Fredrickson-Andersen model actually lends itself to such an analysis. This version of the FA model (and its `bosonic' variant studied in~\cite{garrahanjacklecomtepitardvanduijvendijkvanwijland,garrahan_first-order_2009}) do not possess a finite-dimensional geometrical structure, but still allow to take into account the core features of KCMs by using a kinetic constraint where the transition rates for a given site are proportional to the \textit{number} of its occupied neighbors.

In the theory of finite-size corrections in equilibrium first-order transitions is rather recent, and it can be found described in detail in the papers by Borgs and Koteck\'y~\cite{borgskotecky,PhysRevLett.68.1734}, though earlier descriptions exist (in particular for the Ising ferromagnet~\cite{privmanfisher}). One of the key messages to be drawn from these works is that the partition function splits in a sum of individual partition functions for each of the pure phases, with the transition occurring when these weights are equal. One may wonder whether this picture applies, in some sense, to our dynamical phase transitions, while there is no corresponding static partition function let alone any static free energy. We would like to give an argument that points in that direction. Let us explain out our reasoning. When studying temporal large deviation properties of systems endowed with otherwise 
equilibrium Markov dynamics, one is left with studying the spectrum of a modified evolution operator which is no longer stochastic. The latter does not conserve probability anymore, for instance. Yet it can nevertheless be symmetrized by means of the standard Darboux transformation~\cite{vankampen2007spp}, thus yielding a Hermitian operator. Studying the dynamical transition is thereby formally identical to studying a quantum first-order transition. By invoking Nelson's trick (as pedagogically described in Parisi's chapter 19~\cite{parisi:1988:sft}), one can map the quantum mechanical problem back onto a genuine classical and reversible stochastic process. The latter satisfies, as it should, probability conservation and detailed balance. After all, by a series of well-defined mathematical manipulations, dynamical transitions, quantum phase transitions and classical equilibrium phase transitions, are formally equivalent, and so, why bother? If the original classical process displays a dynamical transition, the 
corresponding 
quantum 
system will exhibit a quantum transition and the final  classical system as well. However, while the original process and its quantum counterpart are characterized by a smooth dependence in the control parameter driving the phase transition, the resulting final classical process has a built-in singularity (a singular dependence in the control parameter, and long-ranged effective interactions). Whether and how the standard phenomenology of finite-size scaling applies to this rather nonphysical effective equilibrium process is, in our view, an open question that we shall address in the present work. 

%Suggest the temporal analog holds, as used by Jack and others. But temporal analogy does not allow to see through the subtleties of the thermodynamic limit. Cite the example of an oscillator by Whitelam and Garrahan~\cite{doi:10.1021/jp037653x} in which a dynamical transition can be found when biasing trajectories with the potential energy. This shows that it is necessary to look into finite size effects.

Here is how the paper is organized: in Section~\ref{Sec/Definition of the model and settings} and Section~\ref{roughs_c}, we begin by recalling what the FA model is in its mean-field version and we review the notions of large deviations and dynamical phase transition. 
Then, in Section~\ref{AnalyticalHamiDiff}, we derive the expression of the free energy for 
the corresponding classical equilibrium system of the dynamical phase transition point. We also derive the finite-size correction of it with a perturbative approach, which is used for deriving scaling functions in the following sections.
In Section~\ref{Scalingfunction}, we numerically study the scaling functions around the dynamical transition, and then we derive the analytical expressions corresponding to them in Sections~\ref{subsec:var-inf-size}, \ref{analyticalscalingfunctions} and~\ref{Sec/Scaling function of psi(s)}. For deriving the expressions, we use an Ansatz similar to the one used by Borgs and Koteck\'y for obtaining the finite-size scaling properties in equilibrium first-order transitions~\cite{borgskotecky,PhysRevLett.68.1734}.
From Section~\ref{Sec/MFqf/Setup}, in order to connect our work to recent advances in the realm of quantum phase transitions, we analyze the same quantum mean-field ferromagnet as that studied by J\"org {\it et al.}~\cite{0295-5075-89-4-40004} and later by Bapst and Semerjian~\cite{1742-5468-2012-06-P06007} to provide, for the latter, a picture for the finite-size scaling functions close to criticality.
In Section~\ref{Sec/MFqf/Setup} and Section~\ref{Sec/MFqf/ResultsInInfiniteSizeLimit}, we define the model and review some results in the infinite-size limit. In Section~\ref{Sec/MFqf/ResultsFiniteSizeLimit}, by using our Ansatz, we derive the same expression of the scaling functions in the quantum ferromagnet and numerically check the result.
With our Ansatz, we can also investigate the scaling factor of the quantum first order phase transition. In Section~\ref{Sec/MFqf/scaling factor}, by applying the Ansatz to the model, we re-derive the formula for the scaling factor obtained by Bapst and Semerjian in Ref.~\cite{1742-5468-2012-06-P06007}.
Note last that the reader will find a table of notations in Appendix~\ref{app:tableofnotations}.

%This leads us to the definition of a size-dependent critical control parameter that helps us characterize the finite-size rounding-off in the large system-size limit.  In turn, we use the methods developed by these authors to access the gap of our evolution operator.

\section{finite-size scaling for the mean-field FA model}
\subsection{Definition of the model and settings}
\label{Sec/Definition of the model and settings}
Each of the~$L$ sites of a fully connected graph is occupied by $n_i=0$ or 1 particle. A particle may appear at an empty site $i$, with rate $\frac{c}{L} \sum_{j\neq i} n_j$, and the particle on an occupied site $i$ can disappear with rate $\frac{1-c}{L} \sum_{j\neq i} n_j$. The parameter~$c$ takes values between 0 and 1. In those rates,
$\frac{1}{L}\sum_{j\neq i} n_j$ represents the number of occupied neighbors to the site $i$ (divided by $L$); it is the spin-facilitating factor, encoding the kinetic constraint. Because the system is a mean-field model, transition rate can be written by using only the total occupation number $n\equiv \sum_{i=1}^{L} n_i$. The transition rate for $n \rightarrow n\pm 1$ is
the sum of the transition rates for each site. That is,
\begin{equation}
w(n\rightarrow n+1)=\sum_i (1-n_i) \frac{c}{L}\sum_{j\neq i} n_j=\frac 1L cn(L-n)
\label{TransitionRateDefinition}
\end{equation}
where $n$ is the kinetic constraint and $L-n$ enumerates the number of empty sites potentially subjected to a creation; and
\begin{equation}
w(n\rightarrow n-1 )=\sum_{i}n_i \frac{1-c}{L}\sum_{j\neq i} n_j =\frac 1L (1-c)n(n-1)\,.
\end{equation}
where $n-1$ is the kinetic constraint and $n$ enumerates the number of occupied sites potentially subjected to an annihilation.
The escape rate is written as
\begin{equation}
r(n)=\sum_{n^{\prime}}w(n\rightarrow n^{\prime})=cn(1-n/L)+(1-c)n(n-1)/L.
\label{escapedefine}
\end{equation}
By convention, we impose $w(n\to n)=0$.
The dynamics for $n$ satisfies the detailed balance condition 
\begin{equation}
  P_{\rm eq}(n)w(n\to n') =  P_{\rm eq}(n')w(n'\to n) \quad (\forall\: n,n')
  \label{eq:DBPeq}
\end{equation}
with respect to the equilibrium distribution function $P_{\rm eq}(n)$ 
\begin{equation}
P_{\rm eq}(n) = \frac{L!}{n! (L-n)!}\frac{1}{1-(1-c)^L}c^n (1-c)^{L-n}.
\label{eq:defPeq}
\end{equation}
Note that we completely omit the $n=0$ state because the system never reaches $n=0$ whenever the initial distribution function has zero probability for the $n=0$ state (this state is isolated). The model is thus described by $1\leq n\leq L$. The expectation value of $n$ is $\left \langle n \right \rangle_{\rm eq}=cL+O((1-c)^L)$, the variance of $n$ is $\left \langle (n-\left \langle n \right \rangle)^2\right \rangle_{\rm eq}=c(1-c)L+O((1-c)^L)$ and the expectation value of the escape rate is $\left \langle r \right \rangle_{\rm eq}=2(1-c)c^2L+O(1)$. The parameter $c$ thus represents the mean density of occupied sites, in the large size limit. We also note that there is a large deviation principle for the probability distribution of the fraction $\rho =n/L$ of occupied sites. The distribution function 
$L^{-1}P_{\rm eq}(L\rho)$
for $\rho$ has an asymptotic expression $L^{-1}P_{\rm eq}(L\rho)\sim \ee^{-Lf_\eee(\rho)}$ at large~$L$, where $f_\eee(\rho)$ is the  large deviation function, and is equal to
\begin{equation}\label{eqfreef}
f_\eee(\rho) = (1-\rho)\log\frac{1-\rho}{1-c} + \rho \log \frac{\rho}{c}.
\label{eq:refIequilib}
\end{equation}
We will loosely adopt the {\it free energy} terminology when speaking about the particle number large deviation $f_\eee(\rho)$ (this applies also to the rest of the paper). The expression in~\eqref{eqfreef} contains an entropic contribution only.

Given a time-interval $[0,t]$, the activity $K_t$ is a trajectory-dependent observable defined as the total  
number of configuration changes the system has undergone up until time $t$. By using the activity, the free energy in the sense of thermodynamic formalism is defined as
\begin{equation}
\psi (s) = \frac{1}{L}\lim_{t\rightarrow \infty} \frac{1}{t} \log \left \langle \ee^{-sK_t} \right \rangle,
\label{eq:defpsis}
\end{equation}
where the average is done over trajectories of duration $t$. {In order to distinguish $\psi(s)$ from the equilibrium free energy $f_{\rm e}(\rho)$, hereafter, we call $\psi(s)$ dynamical free energy.} By considering an appropriately biased dynamics, one can show~\cite{derrida_exact_1998,lecomte_thermodynamic_2007} that the {dynamical} free energy $\psi(s)$ is determined by the largest eigenvalue problem
\begin{equation}
\sum_{n}\Phi_{\rm L}(n) \evol_{n,n^{\prime}}=L\psi(s) \Phi_{\rm L}(n'),
\label{largesteigenvalue}
\end{equation}
where $\evol$ is a $L\times L$ matrix of entries
\begin{equation}
\evol_{n,n^{\prime}}=w(n^{\prime}\rightarrow n)\ee^{-s}-\delta_{n,n^{\prime}}
r(n)
\label{defmatrixL}
\end{equation}
and $\Phi_{\rm L}(n)$ is the left eigenvector corresponding to the largest eigenvalue $L\psi(s)$ of the matrix of entries $\evol_{n,n'}$. 
Note that the first term of~\eqref{defmatrixL} is purely non-diagonal and corresponds to the actual changes of state in the system, while the last term is purely diagonal and corresponds to the escape rate of configurations.
By symmetrizing the matrix, we obtain a mapping to a quantum eigenenergy problem as announced in the {\it Introduction}. Indeed, if we define a matrix $W^\sym$ as 
$W^\sym_{n,n^{\prime}}=\evol_{n,n^{\prime}}(P_{\rm eq}(n^{\prime})/P_{\rm eq}(n))^{1/2}$,
% checked ok
it is easy to show that $W^\sym$ is symmetric, by using the detailed balance condition~\eref{eq:DBPeq}. Furthermore, from the definition itself, we have a relationship between $\Phi_{\rm L}(n)$ and the largest eigenvector $\Phi(n)$ of the symmetrized matrix $W^\sym$ as 
$\Phi_{\rm L}(n) P_{\rm eq}(n)^{1/2}=\Phi(n)$.
% checked ok
Thus, the problems of diagonalizing the matrices $W$ and $W^\sym$ are equivalent. For the ground state energy of a quantum state, a variational principle is well known. By applying it to our case, we obtain a variational expression for the {dynamical}
free energy $\psi(s)$ as
\begin{equation}
L \psi (s) = \max _{\Phi^{\tomax}>0} \frac{\sum_{n,n^{\prime}}\Phi^{\tomax}(n) W^\sym_{n,n^{\prime}} \Phi^{\tomax}(n^{\prime}) }{\sum_{n}\Phi^{\tomax}(n)^2}.
\label{variationalprinciple}
\end{equation}
For a generic symmetric matrix $W^\sym$, the maximization principle is over non-zero vectors.
Here, the optimal vector $\Phi^{\tomax}$ is unique up to a multiplicative factor, and equal to $\Phi$, whose components are strictly positive. This allows us to restrict the maximization to vectors of strictly positive components as in~\eref{variationalprinciple}, without loss of generality.

We denote by $K(s)$ the derivative of $-\psi(s)$ with respect to $s$. 
From the definition~\eref{eq:defpsis} of $\psi(s)$, one has
\begin{equation}
  K(s) = \lim_{t\to\infty} \frac 1 {Lt} \frac{\langle K_t \ee^{-s K_t}\rangle}{\langle \ee^{-s K_t}\rangle}=-\frac{\dd\psi}{\dd s}.
  \label{eq:identitypsiprimeKs}
\end{equation}
The observable $K(s)$ thus describes, in the large-time limit, the average activity of trajectories followed by the system, biased towards either active ($s<0$) or inactive ($s>0$) regions of the space of possible trajectories.

For several classes of KCMs (including the mean-field FA model considered here), $K(s)$ displays a first-order phase transition in the large system-size limit~\cite{garrahanjacklecomtepitardvanduijvendijkvanwijland,garrahan_first-order_2009}: 
\begin{equation}
  \lim_{L\to\infty} K(s) 
  \cases{
    >0 & for \quad $s\leq 0$ (active phase)\\ 
    =0 &for \quad $s> 0$ (inactive phase)\\
  }
\end{equation}
The non-biased steady state $s=0$ lies at the coexistence between the two dynamical phases, characterized by extensive ($K_t=O(L)$) and sub-extensive ($K_t=O(L^0)$) values of the activity in the active and inactive regime respectively.
In contrast to from the equilibrium statistical physics, the parameter $s$ is not a physical field that can be tuned to induce the phase transition --~in the same way as the magnetic field in the Ising model. 
Indeed, the dynamics directly described by the biased evolution operator~\eqref{defmatrixL} does not preserve probability and, to be interpreted, requires for instance to implement a population dynamics picture~\cite{giardina_direct_2006,lecomte_numerical_2007,giardina_simulating_2011}.

However by defining appropriate transition rates, one can build a probability-preserving stochastic dynamics yielding $K(s)$, as extensively discussed in Refs~\cite{nemoto_variational_2011,nemoto_thermodynamic_2011} (see also~\cite{jack_large_2010}). The result is
\begin{equation}
K(s) = \lim_{t\to\infty} \frac 1{Lt} \left \langle K_{t} \right \rangle_{\rm st}^{s},
\end{equation}
where $\left \langle\cdot \right \rangle_{\rm st}^{s}$ is the stationary average for the dynamics defined by the modified transition rates
% $w_{\rm s}(n\rightarrow n^{\prime})$ 
\begin{equation}
w_{s}(n\rightarrow n^{\prime}) = w(n\rightarrow n^{\prime})\frac{\Phi_{\rm L}(n^{\prime})}{\Phi_{\rm L}(n)}\ee^{-s}.
\label{eq:defws}
\end{equation}
The modified dynamics also satisfies detailed balance provided the original one does, with respect to the modified equilibrium distribution
\begin{equation}
P^{s}(n)=C P_{\rm eq}(n) \Phi_{\rm L}(n)^2\,.
\label{distributionfunction_s}
\end{equation}
Here $C$ is the normalization constant. 
The advantage of the modified dynamics is that it preserves probability, but the price to pay is that it involves the left eigenvector~$\Phi_{\rm L}$, difficult to obtain in general if one wants to render the modified rates explicit.

At $s=0$, the property $\sum_{n'}W_{n'n}=0$ ensuring conservation of probability, together with Perron-Frobenius theorem, ensures that $\Phi_{\rm L}(n)=1$ is the unique left eigenvector of the matrix $W|_{s=0}$, of maximal eigenvalue 0. At $s\neq 0$, the maximal left eigenvector $\Phi_{\rm L}$ takes a less simple form.
The free energy is defined as $-\log P^{s}(n)$ and the free energy difference~\cite{nemoto_variational_2011,nemoto_thermodynamic_2011} as
\begin{equation}
\Delta F_s(n) =-2\log \Phi_{\rm L}(n).
\label{HamiDiff_def}
\end{equation}
It encodes (see~\eqref{distributionfunction_s}) the modification brought by~$s$ to the $s=0$ equilibrium state $P_{\rm eq}$. We indeed have $\Delta F_{s=0}(n)=0$.

We now turn onto the study of the dynamical phase transition presented by our model of interest, and to the study of its finite-size scaling.
In subsection~\ref{roughs_c}, we focus on the critical point $s_c=s_c(L)$ separating active and inactive dynamical phases, by studying its behavior as $L$ becomes large.
It turns out that not only the behavior of the largest eigenvalue, but also of its corresponding eigenvector is essential to the understanding of the finite-size corrections. In subsection~\ref{AnalyticalHamiDiff} we focus on exact point of phase coexistence $s=s_c$, and subsection~\ref{Scalingfunction} we determine the generic form of the eigenvector in the vicinity $s\approx s_c$ of the critical point. 
In subsection~\ref{subsec:var-inf-size}, we examine how a variational principle can be used to determine this form,
and in subsection~\ref{analyticalscalingfunctions} we actually compute the scaling function for the mean density of occupied sites and for its variance.
Last, in subsection~\ref{Sec/Scaling function of psi(s)}, we write down the explicit expression of the scaling functions for the two first
derivatives of the dynamical free energy $\psi(s)$ encoding the mean and the variance
of the activity. We thus fully describe the finite-size behavior of the fluctuation of the
activity in our model.

\subsection{The dynamical free energy and the upper bound for $\lambda_\rc$}
\label{roughs_c}
\begin{figure}
\centering
\includegraphics[width=9cm ]{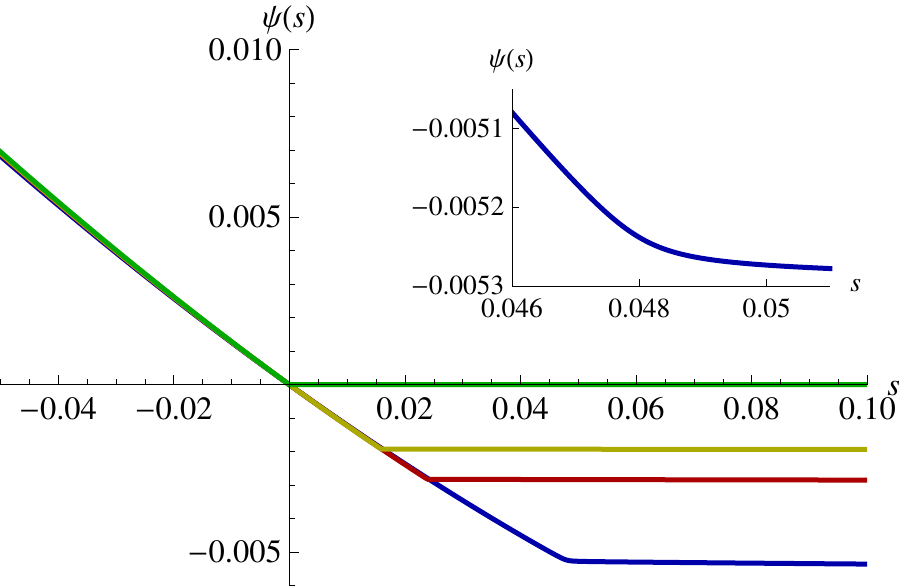}
\caption{\label{DynamicalFreeEnergy}%
Dynamical free energy $\psi(s)$ for $c=0.3$, $L=50$ (\textcolor{darkblue}{blue}), $L=100$ (\textcolor{darkred}{red}), $L=150$ (\textcolor{darkyellow}{yellow}) and $L=\infty$ (\textcolor{darkgreen}{green}).
By locating numerically the maximum of $\psi''(s)$, we estimate that $s_\rc$ is equal to $0.0479...$ for $L=50$, $0.02390...$ for $L=100$ and $0.01591...$ for $L=150$. The value is close to $1/(2Lc(1-c))$, which takes
$0.04762...$, $0.02381...$ and $0.01587...$ for $L=50,100$, and $150$, respectively.
The $L=\infty$ curve is obtained from the variational formula~(\ref{variationalprinciple5}).
{The inset illustrates the rounding of the cusp due to finite-size, for $L=50$, in the region of the transition.}
}
\end{figure}

We present in Fig.~\ref{DynamicalFreeEnergy} the numerical evaluation of the dynamical free energy $\psi(s)$ for $c=0.3, L=50,100,150$, obtained by solving the largest eigenvalue problem~(\ref{largesteigenvalue}) numerically.
We observe a remarkable point in each plot where $\psi(s)$ presents a cusp, meaning that $\psi''(s)$ is sharply peaked. {We note that the cusp is rounded in the finite-size systems as shown in the inset of Fig.~\ref{DynamicalFreeEnergy}, which becomes sharper as the system size becomes larger.} We denote the corresponding $s$ by $s_\rc$, and call it finite-size critical point: 
\begin{equation}
{s_\rc \equiv {\rm Argmax}~ \psi''(s).}
\label{s_cDefinition}
\end{equation}
As reported in Refs.~\cite{garrahanjacklecomtepitardvanduijvendijkvanwijland,garrahan_first-order_2009}, 
$s_\rc$  deviates from the origin, but goes to zero as $L\to\infty$.

We derive below an upper bound for {a scaled critical point}
\begin{equation}
{\lambda_c=\lim_{L\to\infty} Ls_\rc.}
\label{lambda_cDefinition}
\end{equation}
By taking $s\rightarrow \infty$ in~(\ref{largesteigenvalue}), we obtain the asymptotic behavior of $\psi (s)$ at large $s$
\begin{equation}
\lim_{s\rightarrow \infty}\psi (s)=-\frac{1}{L}\min_{n>0}r (n)=-c(L-1)/L^2.
\end{equation}
Remarking now from~\eqref{eq:identitypsiprimeKs} that $\psi(s)$ is strictly decreasing, we obtain the inequality
\begin{equation}
\psi(s)>-c(L-1)/L^2.
\label{inequality}
\end{equation}
%
%  strictly convex thanks to the identity
% $\psi''(s)=\lim_{t\to\infty} [\langle K_t^2\rangle_s-{\langle K_t\rangle_s}^2]/t$
%
On the other hand, by expanding $\psi(s)$ around $s=0$, we obtain that
\begin{equation}
\psi(s)=-\left \langle r \right \rangle_{\rm eq} \frac{s}{L}+O(s^2)=-2(1-c)c^2s+O(s^2).
\label{eq:expansionpsismalls}
\end{equation}
Note that this expansion is valid only in the region around $\lambda=0$ where $\psi(s)$ remains analytic in the large size limit. 
To identify this region and define properly its scaling, let us now introduce the rescaled {dynamical} free energy~\cite{bodineautoninelli,bodineaulecomtetoninelli}
\begin{equation}
  \varphi_L(\lambda)=L \psi(\lambda/L)
  \label{rescaledFreeEnergy}
\end{equation}
The transition point $s_\rc$ now defines a critical value $\lambda_\rc=\lim_{L\to\infty} L s_c$ where $\varphi_\infty(\lambda)=\lim_{L\to\infty}\varphi_L(\lambda)$ presents a non-analyticity. The previous expansion~\eref{eq:expansionpsismalls} becomes
\begin{equation}
  \varphi_L(\lambda)=-2(1-c)c^2 \lambda+O(\lambda^2/L).
\end{equation}
Thus one has $\varphi_\infty(\lambda)=-2(1-c)c^2 \lambda$ in the region around $\lambda=0$ where $\varphi_\infty(\lambda)$ is analytic,
while the inequality~\eref{inequality} yields $\varphi_\infty(\lambda)\geq -c$ for all values of $\lambda$.
This implies that $\varphi_\infty(\lambda)$ becomes non-analytic at a point $\lambda_\rc$ bounded as
\begin{equation}\label{criticlam}
\lambda_\rc \leq \frac{1}{2c(1-c)}
\end{equation}
Interestingly, this expression also gives a good approximation for~$s_\rc$: The behavior $s_\rc\simeq 1/(2Lc(1-c))$ can be checked numerically as displayed in Fig.~\ref{DynamicalFreeEnergy}. In Appendix~\ref{app:sc_expression_derivation}, building on our results, we shall derive the equality $s_\rc = 1/(2Lc(1-c))+O(1/L^2)$.

Remark last that in finite-dimensional FA models, such an approach also yields an upper bound for $\lambda_\rc$ which however is not the value of $\lambda_\rc$, due to the complex interfacial spatial structure of the steady state around $\lambda_\rc$~\cite{bodineautoninelli,bodineaulecomtetoninelli}.

\subsection{The free energy difference at $s=s_\rc$ and the finite-size correction}
\label{AnalyticalHamiDiff}
\begin{figure}
\centering
\includegraphics[width=9cm ]{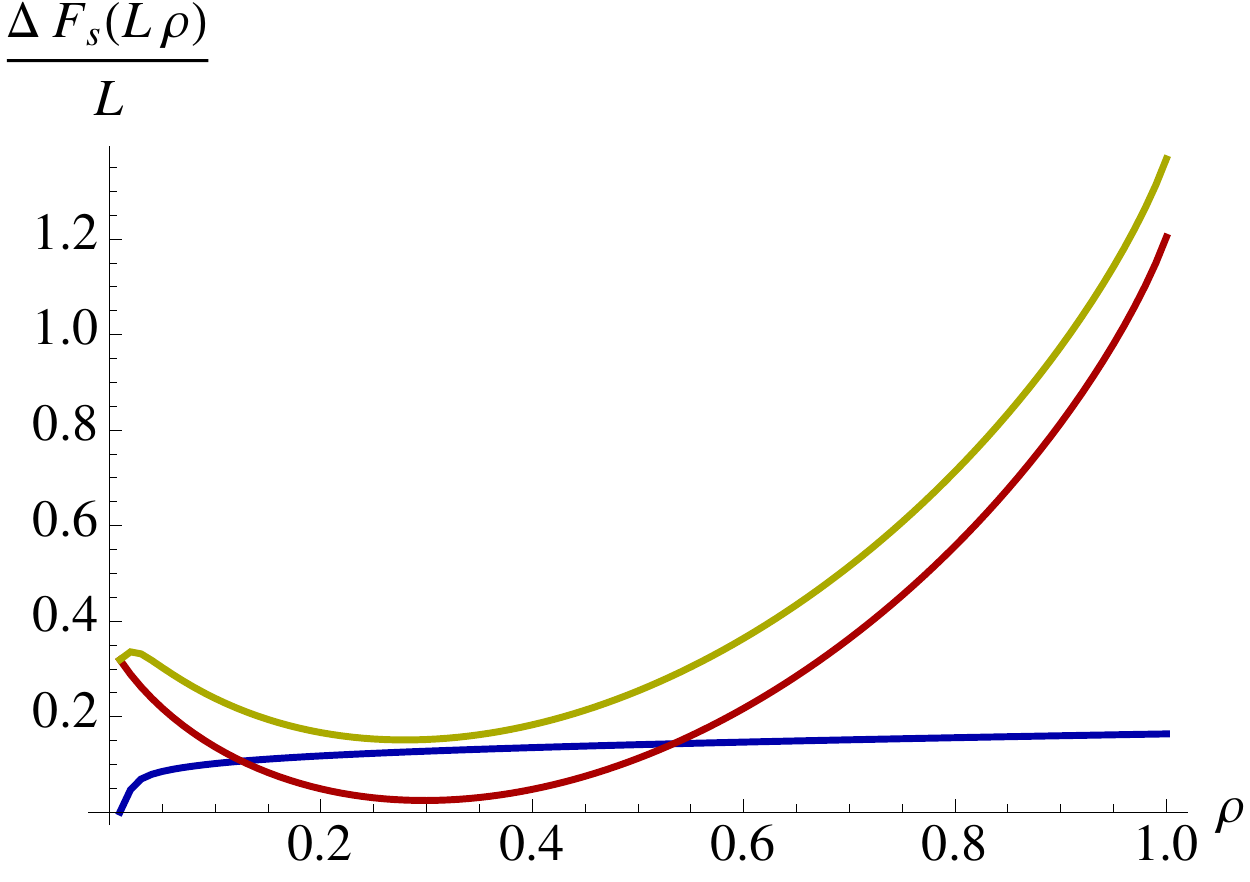}
\includegraphics[width=9cm ]{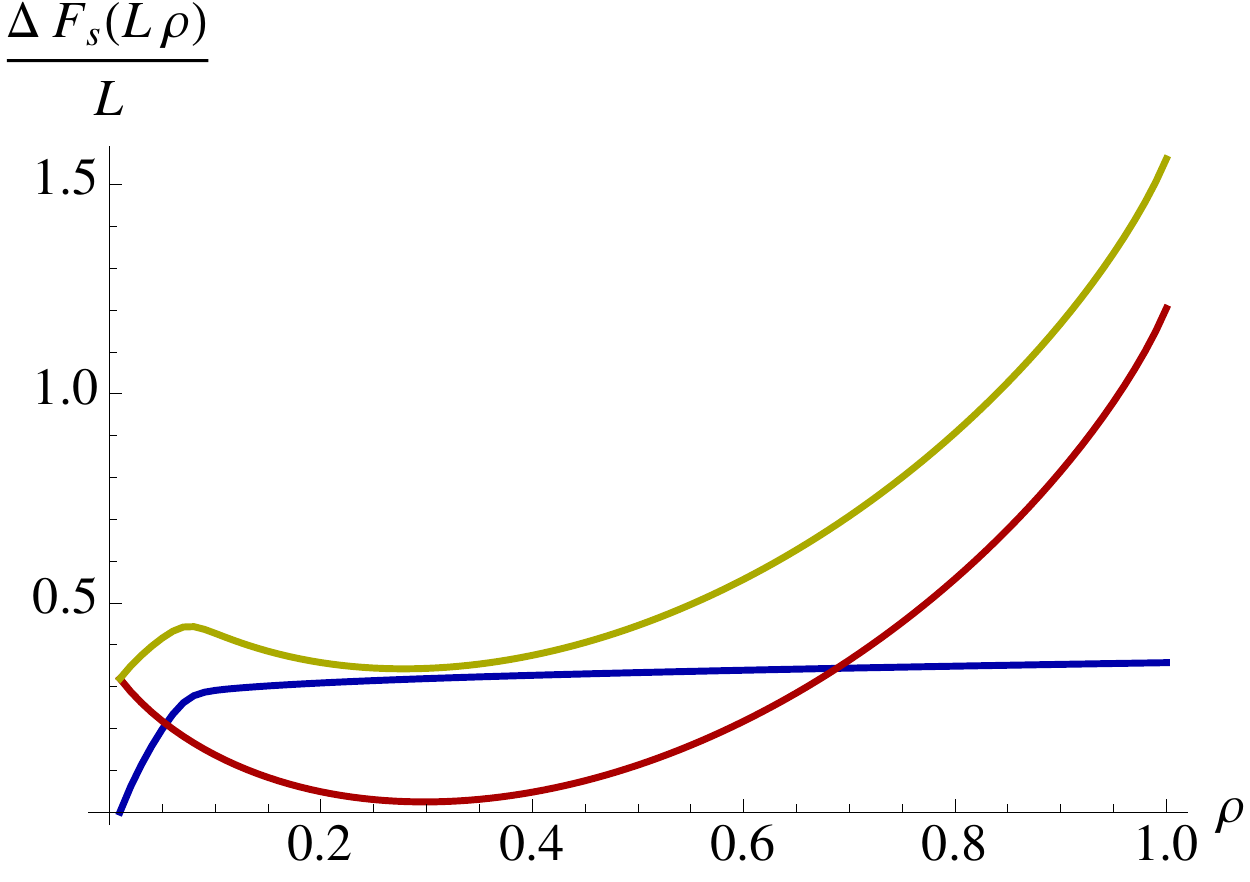}
\includegraphics[width=9cm ]{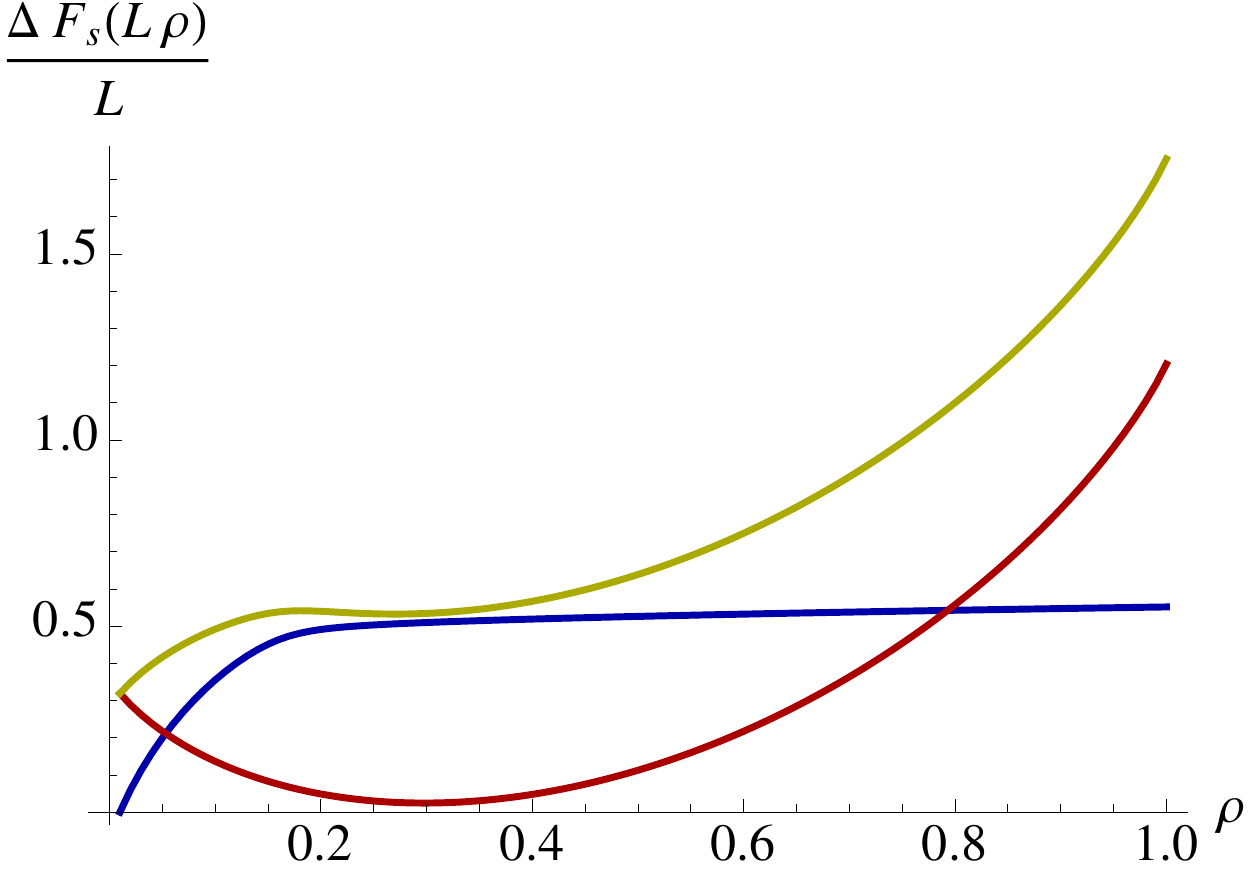}
\caption{\label{Hamiltoniandifference}The free energy difference $\Delta F_{s}(\rho L)/L$ (\textcolor{darkblue}{blue}), the original free energy $-\log P_{\rm eq}(\rho L)/L$ (\textcolor{darkred}{red}) and the modified one $-\log P^{s}(\rho L)/L=-\log P_{\rm eq}(\rho L)/L + \Delta F_{s}(\rho L)/L$ (\textcolor{darkyellow}{yellow}) for $c=0.3, L=100$. We set $s=0.95 s_\rc$ (top), $s=s_\rc$ (middle) and $s=1.05 s_\rc$ (bottom). 
In the $L\to\infty$ limit, those functions converge to analytical forms described by~\eqref{largedeviation_modifiedhamiltonian1}, \eqref{largedeviation_modifiedhamiltonian2} and~\eqref{equationForrhoc}, depicted in Fig.~\ref{FigLargedeviation_sc}.
}
\end{figure}

Let us now study numerical examples of the free energy difference for ${s=0.95s_\rc}$, $s=s_\rc$ and $s=1.05\,s_\rc$: in each subfigure of Fig.~\ref{Hamiltoniandifference} are displayed the free energy difference $\Delta F_{s}(\rho L)/L$, the original free energy $-\log P_{\rm eq}(\rho L)/L$, and the modified one $-\log P^{s}(\rho L)/L$ as functions of the density $\rho$.
For $s=s_\rc$, one observes that the modified free energy reaches its minimum value at the two values of the density ${\rho=n/L}$ of occupied sites characterizing the inactive  ($\rho \simeq 0$)  and the active ($\rho \simeq c$) configurations. This indicates that the first order phase transition observed along direction $s$ at $s=s_\rc$ also reflects itself in a first order coexistence in the density $\rho$: precisely at $s=s_\rc$, the two competing phases have the same weight. 
In this subsection, we consider the analytical expression of the free energy (difference).
First, we derive the analytical expression of the free energy $f_{s}(\rho)\equiv - \lim_{L\rightarrow \infty} \frac{1}{L}\log P^s_{\rm eq}(\rho L)/L $ at $s=s_c$, which describes the distribution of particle occupation at the transition point in the infinite-size limit.
Then, we derive the finite-size correction of these expressions, which will be used for deriving the
scaling behavior of this first order phase transition in the next section.

\begin{figure}[t]
\centering
\includegraphics[width=8cm ]{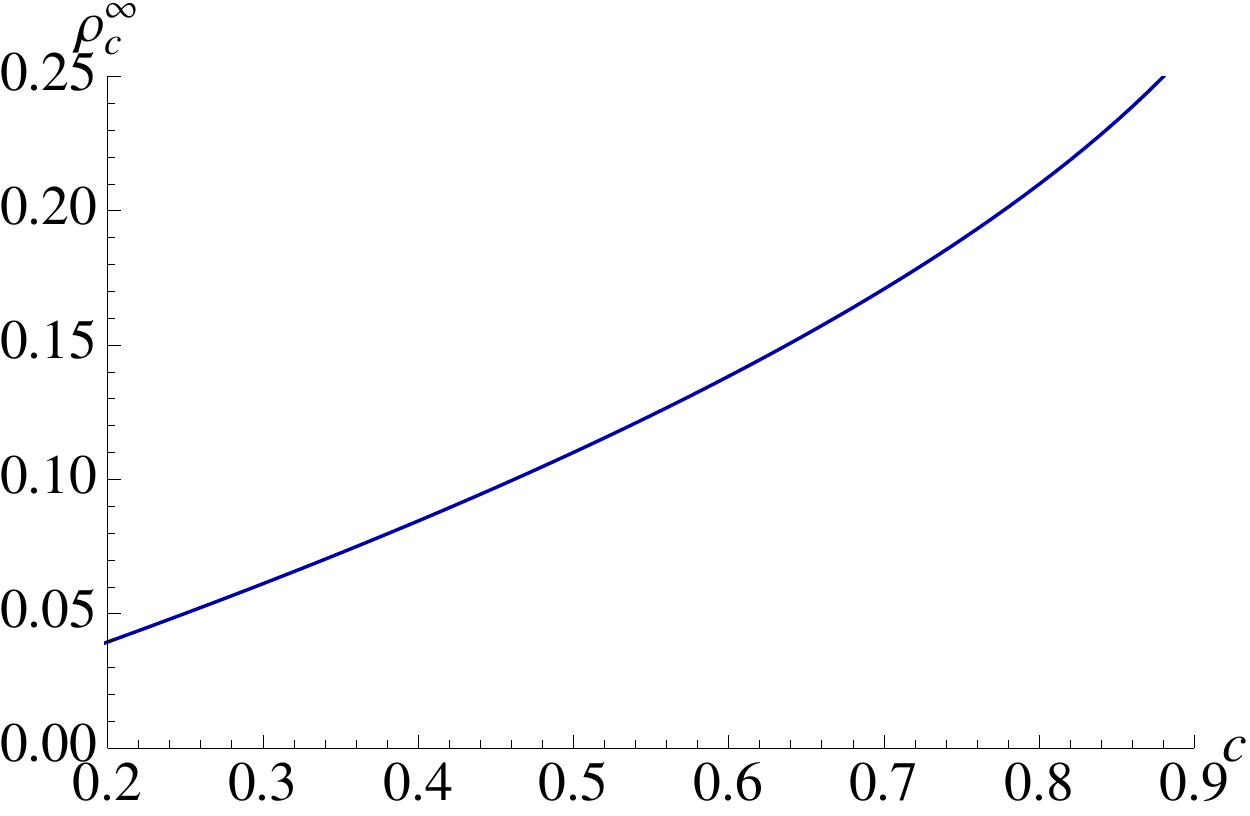}
\caption{\label{FigRhocVsc}%
Connecting point $\rho_{\rm c}^\infty$ in the infinite-size limit, as a function of $c$, obtained from solving~(\ref{equationForrhoc}).
}
\end{figure}

\begin{figure}[t]
\centering
\includegraphics[width=8cm ]{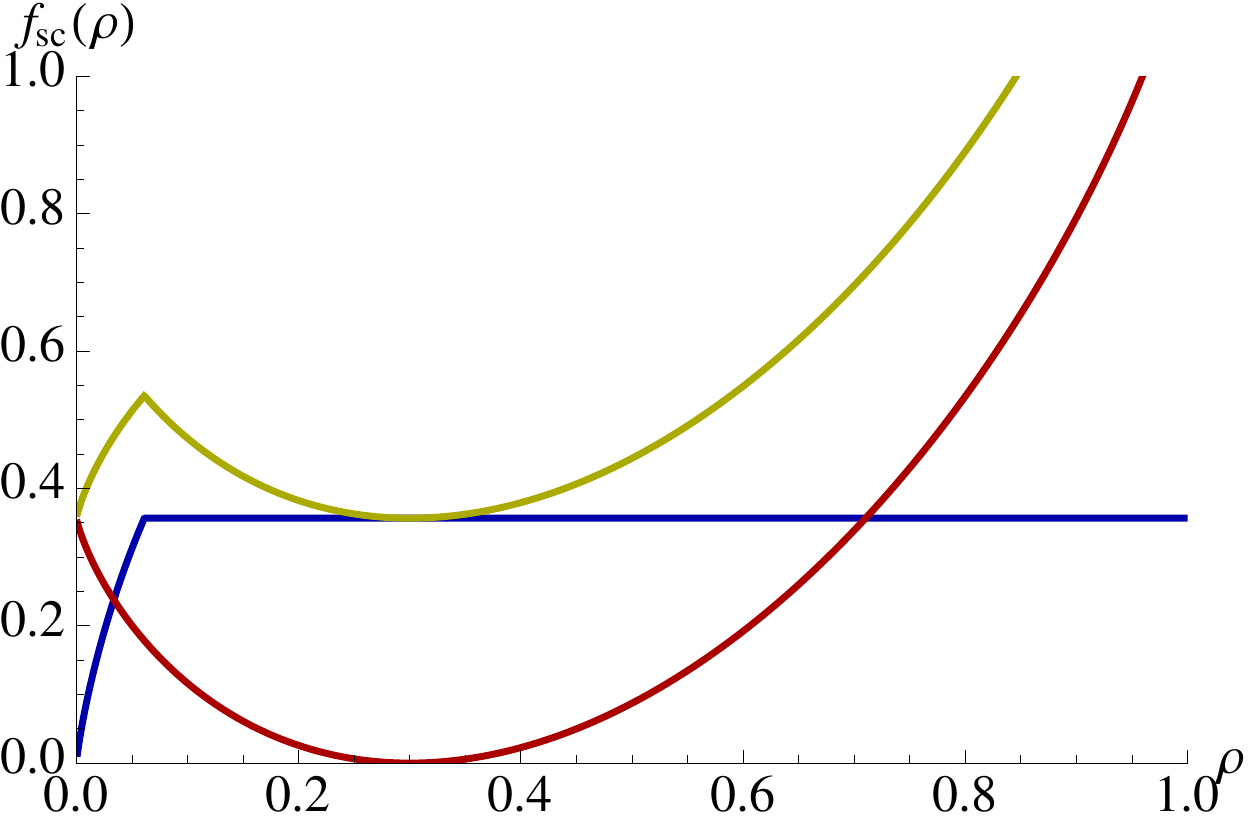}
\caption{\label{FigLargedeviation_sc}%
The free energy 
$f_{\rm e}(\rho)+\Delta f_{\rm s_\rc}(\rho)$ (\textcolor{darkyellow}{yellow}) for $s=s_c$ and the modifying free energy
$\Delta f_{\rm s_\rc}(\rho)$ (\textcolor{darkblue}{blue}) for $s=s_c$ and the unbiased free energy $f_{\rm e}(\rho)$ (\textcolor{darkred}{red}).
The point $\rho=\rho_{\rm c}^{\infty}$ of non-analyticity of the two first functions is provided by the solution of equation~\eqref{equationForrhoc}.
}
\end{figure}

We start with the eigenvalue equation~(\ref{largesteigenvalue}). 
Since we focus on $s=s_c$,
we can set
\begin{equation}
L\psi(s_\rc)=-c+O(1/L),
\end{equation}
which leads to
\begin{eqnarray}
& \Phi_{\rm L}(n+1)\frac{cn}{L}(L-n)\ee^{-s_\rc}+\Phi_{\rm L}(n-1)\frac{(1-c)}{L}n(n-1)\ee^{-s_\rc}
\nonumber \\
& -\Phi_{\rm L}(n)\left [c\frac{n}{L}(L-n)+(1-c)\frac{n}{L}(n-1) -c +O(1/L) \right ] =0.
\label{EigenEquation_Derivation}
\end{eqnarray}
Now, we assume the large deviation principle of $\Phi_{\rm L}(n)$, 
\begin{equation}
{ \Phi_{\rm L}(n)=\ee^{-L\Delta f_{\rm s_\rc}(n/L)/2},}
\label{DensityOfFreeEnergyDifference}
\end{equation}
which defines $\Delta f_{\rm s_\rc}$. By substituting $\Phi_{\rm L}(n)$ in~(\ref{EigenEquation_Derivation}) by the large deviation expression and evaluating the leading order with an assumption of differentiability of $\Delta f_{\rm s_\rc}(\rho)$, we obtain an equation for determining $\Delta f_{\rm s_\rc}(\rho)$. That is, 
\begin{eqnarray}
& \ee^{-(1/2)\partial \Delta f_{\rm s_\rc}(\rho)/\partial \rho}c\rho (1-\rho)
+\ee^{(1/2)\partial \Delta f_{\rm s_\rc}(\rho)/\partial \rho}(1-c)\rho^2 
\nonumber \\
& -\left [c\rho (1-\rho)+(1-c)\rho^2\right ] =0.
\end{eqnarray}
By solving this equation, we obtain two expressions of $\partial \Delta f_{\rm s_\rc}(\rho)/\partial \rho$ as
\begin{equation}
\partial \Delta f_{\rm s_\rc}(\rho)/\partial \rho=0
\end{equation}
and
\begin{equation}
\partial \Delta f_{\rm s_\rc}(\rho)/\partial \rho=-2\log\left [\frac{(1-c)\rho}{c(1-\rho)} \right ] = -2\frac{\partial f_{\rm e}(\rho)}{\partial \rho},
\end{equation}
which leads to
\begin{equation}
\Delta f_{\rm s_\rc}(\rho)=\rm const.
\end{equation}
and
\begin{equation}
\Delta f_{\rm s_\rc}(\rho)=-2f_{\rm e}(\rho) + \rm const.
\end{equation}
We connect these two functions. We call the connecting point $\rho_{\rm c}^{\infty}$.
By referring the numerical result in Fig.~\ref{Hamiltoniandifference}, we conjecture that $\Delta f_{\rm s_\rc}(\rho)$ becomes
\begin{equation}
\Delta f_{\rm s_\rc}(\rho)=-2f_{\rm e}(\rho) + 2f_{\rm e}(0).
\label{largedeviation_modifiedhamiltonian1}
\end{equation}
for $\rho \leq \rho_c^{\infty}$ and
\begin{equation}
\Delta f_{\rm s_\rc}(\rho)=-2f_{\rm e}(\rho_{\rm c}^{\infty}) + 2f_{\rm e}(0).
\label{largedeviation_modifiedhamiltonian2}
\end{equation}
for $\rho > \rho_{\rm c}^{\infty}$. Here, $\rho_{\rm c}^{\infty}$ is determined from the condition of first order phase transition
expressing here that the inactive  ($\rho \simeq 0$)  and the active ($\rho \simeq c$) configurations have the same weight
\begin{equation}
f_{\rm e}(c) + \Delta f_{\rm s_\rc}(c)=f_{\rm e}(0) + \Delta f_{\rm s_\rc}(0),
\end{equation}
which leads to
\begin{equation}
2\left [
(1-\rho_{\rm c}^{\infty})\log\frac{1-\rho_{\rm c}^{\infty}}{1-c} + \rho_{\rm c}^{\infty}\log \frac{\rho_c^{\infty}}{c}
\right ]
= -\log (1-c).
\label{equationForrhoc}
\end{equation}
In Fig.~\ref{FigRhocVsc}, we plot $\rho_{\rm c}^{\infty}$ as a function of $c$. Also, in Fig.~\ref{FigLargedeviation_sc}, we plot the obtained $\Delta f_{\rm s_\rc}(\rho)$ with
$f_{\rm e}(\rho)+\Delta f_{\rm s_\rc}(\rho)$ and $f_{\rm e}(\rho)$ for $c=0.3$. From the figure, we understand that $\rho_{\rm c}^{\infty}$ is a point of non-analyticity for the free energies: the distribution of the density~$\rho$ of occupied sites in the system is naturally divided into two domains, namely, an active ($\rho>\rho_{\rm c}^{\infty}$) and an inactive ($\rho<\rho_{\rm c}^{\infty}$) domain.

\begin{figure}[t]
\centering
\includegraphics[width=8cm ]{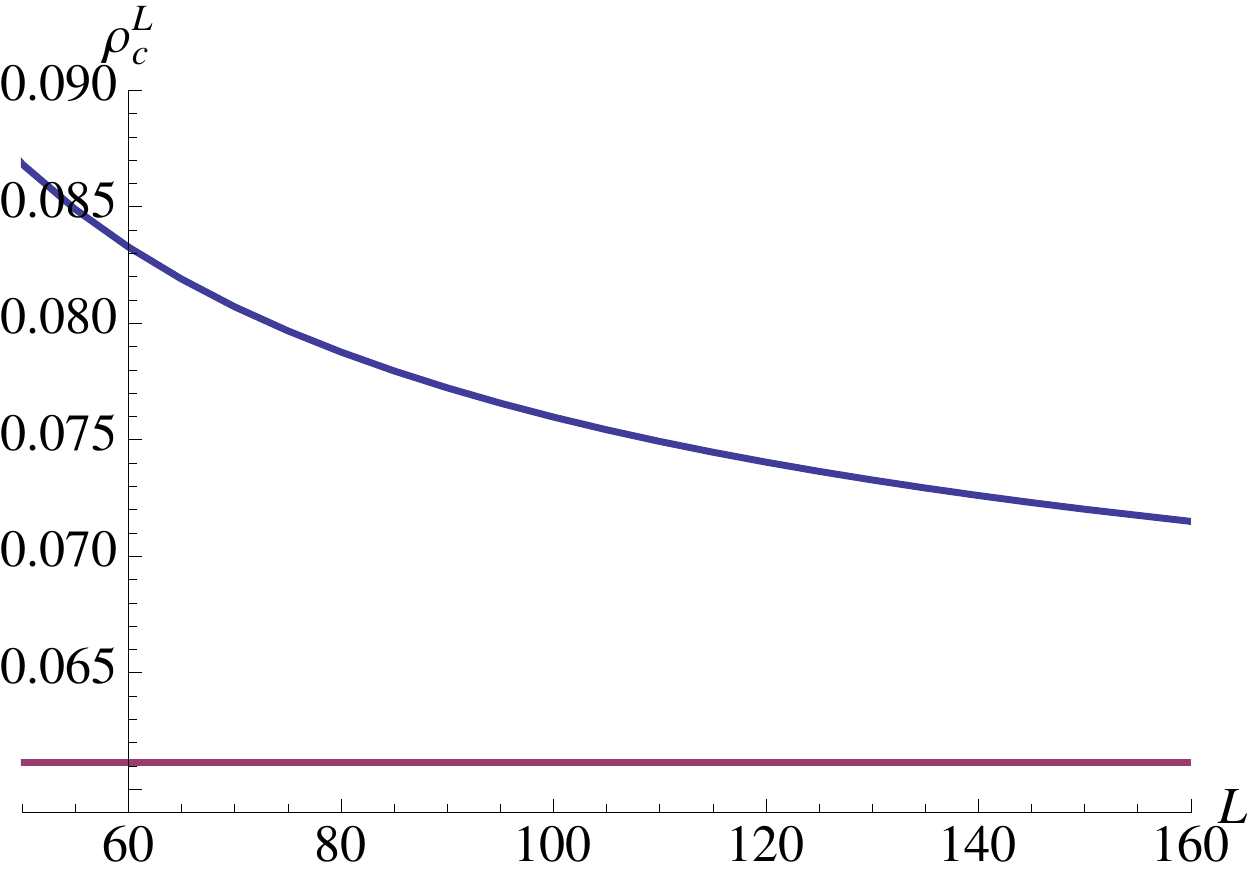}
\caption{\label{FigRhocVsL}%
Numerical example illustrating the importance of finite-size corrections for $\rho_\rc^L$ (in \textcolor{darkblue}{blue}), compared to its infinite-size limit $\rho_\rc^{\infty}$ (in \textcolor{darkpurple}{purple}) for $c=0.3$.
}
\end{figure}

\begin{figure}[t]
\centering
\includegraphics[width=8cm ]{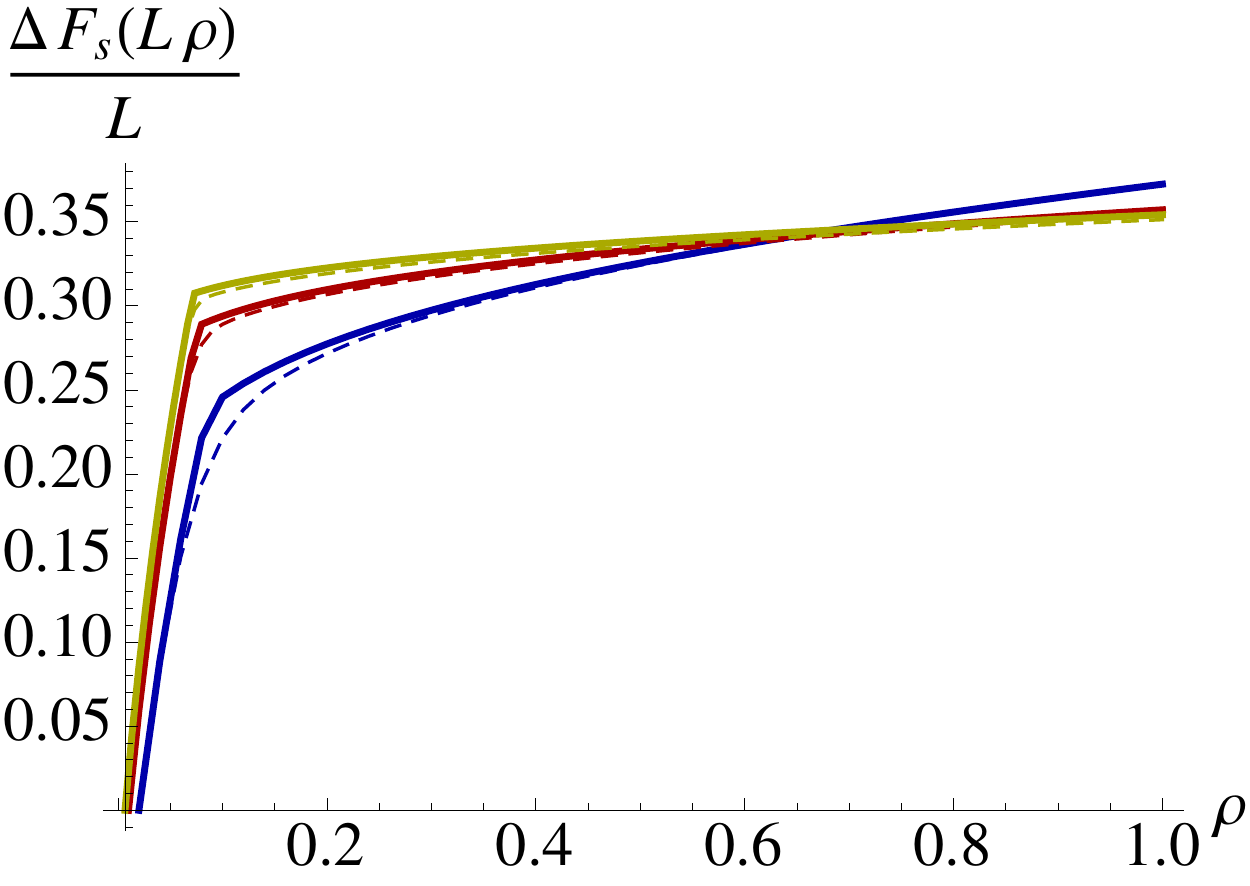}
\caption{\label{Hamiltoniandifference_analy}%
The finite-size correction of the free energy difference for $c=0.3$, $s=s_\rc$, $L=50$ (in \textcolor{darkblue}{blue}), $L=100$ (\textcolor{darkred}{red}) and $L=150$ (\textcolor{darkyellow}{yellow}). We plot analytical finite-size free energy difference obtained from~(\ref{deltafsmall}) and~(\ref{deltaflarge}) as solid lines.
We also plot the  exact numerical results as dashed lines. One observes the emergence of the non-analytic point at $\rho_\rc^{L}$ as $L$ becomes large.
}
\end{figure}

Now, we consider the finite-size correction of the free energy. Those will be crucial to understand the finite-size scaling of the large deviation function, as explained in subsection~\ref{analyticalscalingfunctions}.
For this, we conjecture that the kink-like behavior of $\Delta f_{\rm s_\rc}(\rho)$ appears faster than $O(1/L)$ as suggested by the numerics presented in Fig.~\ref{FigEx_Vari}. From this, even for finite-size systems, we can define the active and the inactive regions. We denote the finite-size connecting point by $\rho_c^{L}$. Precisely, $\rho_{\rm c}^{L}$ is defined as
\begin{equation}
\sum_{n \leq n_\rc^{L}}P^{s}(n) = \sum_{n > n_\rc^{L}}P^{s}(n),  
\label{condition_for_rhoc}
\end{equation} 
with the definition of $n_\rc = \lfloor L \rho_\rc^{L} \rfloor$. Here, as shown in what follows, the distribution function $P^{s}(n)$ itself has a $\rho_c^{L}$ dependence, so that the equation~(\ref{condition_for_rhoc}) can give a non-integer value for $\rho_c^{L}$.
% and where $P^{s}(n)$ is obtained from $\Delta H_{s}(\rho L)/L$ using~\eqref{distributionfunction_s}.
In order to obtain the next order correction of $\Delta f_{\rm s_\rc}(\rho)$ for each active and inactive region, we use a perturbation analysis. That is, for each region, 
we assume the following scaling
\begin{equation}
\Phi_{\rm L}(n)=\exp\left \{
-(L/2) \left [ \Delta f_{s_{\rm c}}(n/L) + (1/L) \Delta f_{s_{\rm c}}^{(1)}(n/L) \right ] \right \}.
\label{Def_Deltaf1}
\end{equation}
Expanding in powers of  $L$ for large $L$, we obtain the correction $\Delta f_{s_{\rm c}}^{(1)}(n/L)$ as
\begin{equation}
\Delta f_{s_{\rm c}}^{(1)}(\rho)=-\log \frac{\rho(1-\rho)}{(c-r)^2} 
- \frac{\rho(1-2c)}{c(1-c)} + \rm const. % - \log \frac{1/L}{1-1/L} + \frac{(1/L)(1-2c)}{c(1-c)} 
\label{deltafsmall}
\end{equation}
for $\rho \leq \rho_{\rm c}^L$, and
\begin{equation}
\Delta f_{s_{\rm c}}^{(1)}(\rho)=-2\left [\frac{\rho(2c - 1)}{2c(1-c)} -  \log\rho\right ] + \rm const.
\label{deltaflarge}
\end{equation}
for $\rho > \rho_c^L$.
See Appendix~\ref{app:Derivation_approximation} for the derivation of these expressions. The two constants in~(\ref{deltafsmall}) and~(\ref{deltaflarge}) are determined from the conditions $\Delta f_{s_{\rm c}}^{(1)}(1/L) =0$, and $\lim_{\rho \rightarrow \rho_{\rm c}^L}\Delta f_{s_{\rm c}}^{(1)}(\rho) = \lim_{\rho \rightarrow \rho_{\rm c}^L} \Delta f_{s_{\rm c}}^{(1)}(\rho)$. We note that the latter constant depends on $\rho_{\rm c}^{L}$. In Fig.~\ref{FigRhocVsL}, we plot numerical examples of $\rho_\rc^L$ obtained from~(\ref{condition_for_rhoc}-\ref{deltaflarge}) as a function of $L$, together with the limiting value~$\rho_\rc^{\infty }$, which is the solution of~(\ref{equationForrhoc}).  
In Fig.~\ref{Hamiltoniandifference_analy}, we compare our finite-size free energy with
the numerical results obtained by direct diagonalization, for $c=0.3$, $s=s_\rc$, $L=50,100,150$. The agreement of both results improves as $L$ increases. 

\subsection{Scaling function around $s=s_\rc$}
\label{Scalingfunction}

As seen in the previous section, the system also displays a phase coexistence in the density $\rho = n/L$ of occupied sites, which plays a role similar to the mean magnetization in the Ising model. For the equilibrium ferromagnet model, the scaling functions describing finite-size scaling around the first order phase transition point have been determined~\cite{borgskotecky,PhysRevLett.68.1734}. 
However, even though our system presents a first order phase transition, it is not trivial to determine such scaling functions because of the asymmetry of the scalings in the active and inactive phases. In what follows, we study the finite-size scaling behavior in our system.

\begin{figure}[h]
\centering
\includegraphics[width=8cm ]{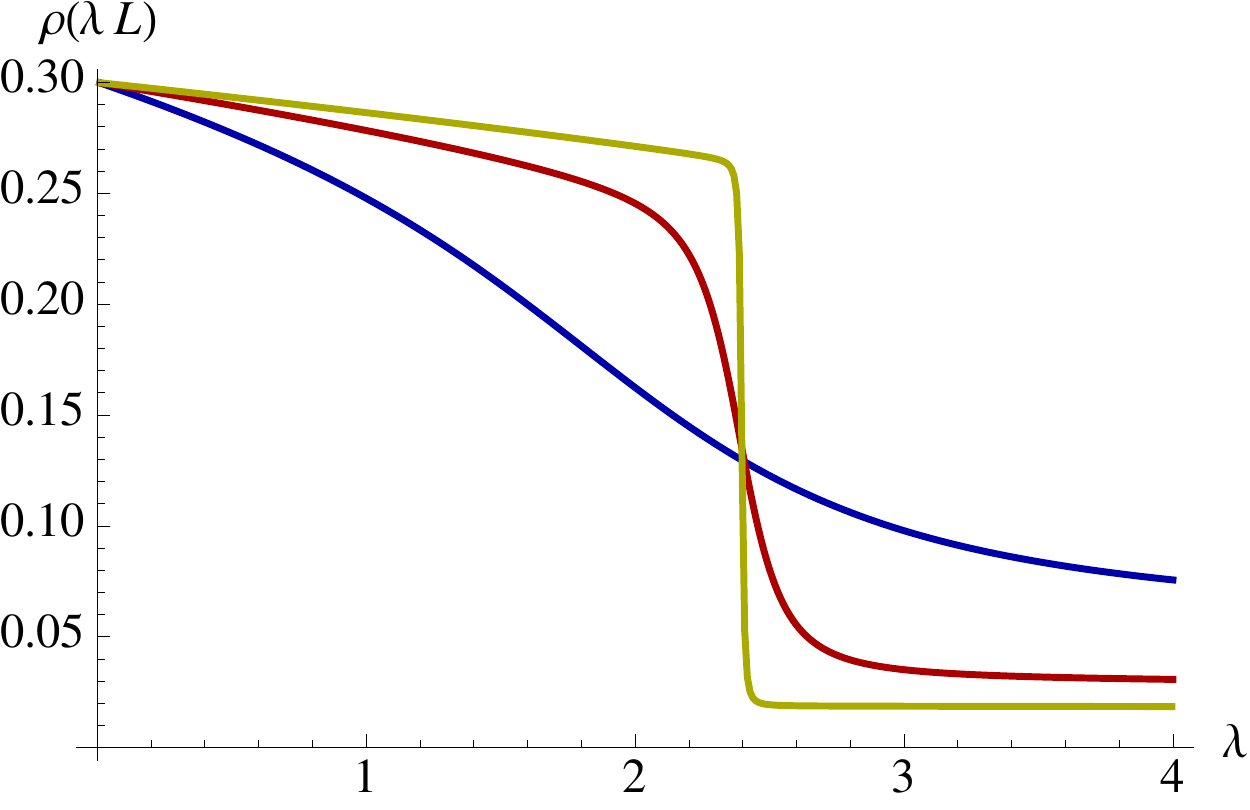}
\includegraphics[width=8cm ]{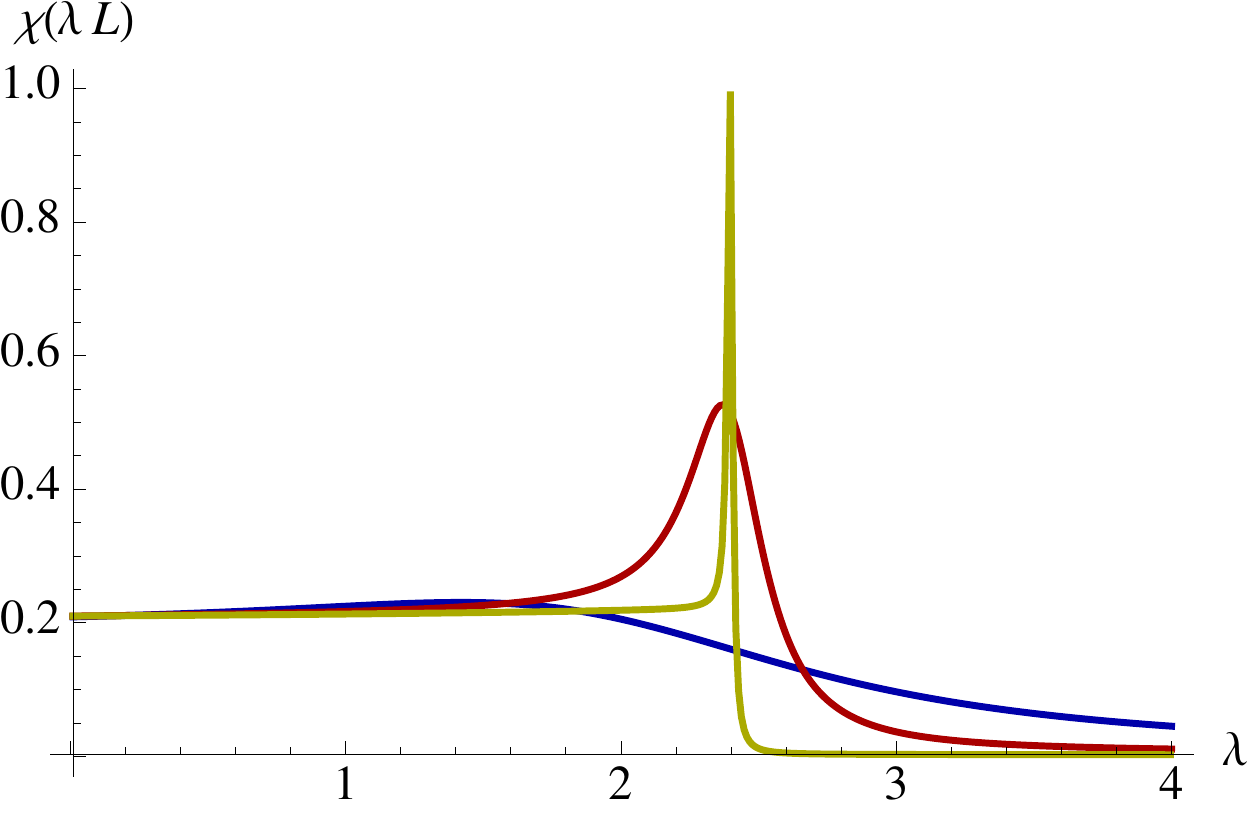}
\caption{\label{FigEx_Vari}%
The expectation value $\rho(\lambda L)$~\eqref{eq:defrhos} (top) and the variance~\eqref{eq:defchis} $\chi(\lambda L)$ (bottom) of the mean occupation $n/L$ in the $s$-modified equilibrium state, plotted as a function of the rescaled variable $\lambda=s/L$. We set $c=0.3$ and $L=20$ (\textcolor{darkblue}{blue}), $L=40$ (\textcolor{darkred}{red}) and $L=60$ (\textcolor{darkyellow}{yellow}). The position of the peak of $\chi(s)$
defines the value of $s_\rc$, which appears on the graphs as a function of $\lambda$ at a value close to the critical $\lambda_\rc=1/(2c(1-c))$ derived in Section~\ref{roughs_c}. For example, the peak occurs approximately at $\lambda=2.398...$ for $L=60$, which is close to the corresponding $\lambda_\rc\simeq 2.381...$}
\end{figure}

\begin{figure}[h]
\centering
\includegraphics[width=9cm ]{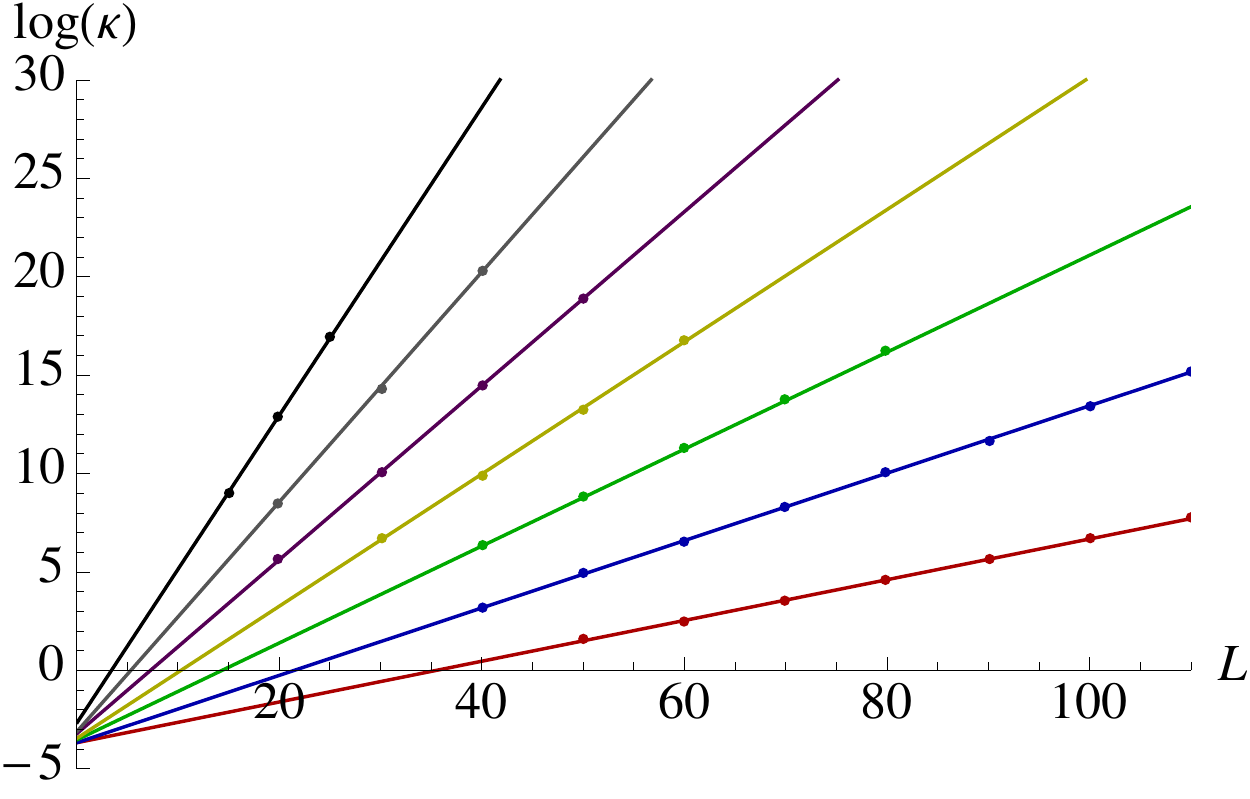}
\caption{\label{ScalingRatio}
The logarithm of the scaling ratio $\kappa$ defined in~\eqref{ScalingRationDef} as a function of~$L$ for $c=0.2$ (\textcolor{darkred}{red}), $c=0.3$ (\textcolor{darkblue}{blue}), $c=0.4$ (\textcolor{darkgreen}{green}),
$c=0.5$ (\textcolor{darkyellow}{yellow}), $c=0.6$ (\textcolor{darkpurple}{purple}), $c=0.7$ (\textcolor{darkgrey}{gray}) and $c=0.8$ (black). Dots were obtained by solving the eigenvalue equation~(\ref{largesteigenvalue}) numerically. Solid lines are linear fits.}
\end{figure}

To do so, we start by considering the statistical properties of the mean occupation $n/L$.
We denote by $\rho(s)$ its expectation value
\begin{equation}
\rho(s) = \sum_{n}(n/L) P^{s}(n),
\label{eq:defrhos}
\end{equation}
and by $\chi(s)$ its variance
\begin{equation}
\chi(s) = L \sum_{n}(n/L - \rho(s))^2 P^{s}(n).
\label{eq:defchis}
\end{equation}
Examples of these functions are shown in Fig.~\ref{FigEx_Vari}, and the latter illustrate how the first order transition materializes around $s=s_\rc$.
One observes in particular that the width of the first order coexistence region shrinks as the system size~$L$ becomes large. 
To estimate this dependence in~$L$, we define a scaling ratio~$\kappa$ as 
\begin{equation} 
\kappa= - \partial \rho(s)/\partial s|_{s=s_\rc},
\label{ScalingRationDef}
\end{equation}
which is inversely proportional to the width of the first order coexistence region, and diverges as $L\to\infty$.  
The logarithm of $\kappa$ for various values $c$ is represented as a function of~$L$ in Fig.~\ref{ScalingRatio}. 
We find that these logarithms are increasing linear functions of~$L$. It indicates that the width of the first order coexistence region (illustrated in Fig.~\ref{FigEx_Vari}) shrinks exponentially with the system size~$L$.

\begin{figure}[h]
\centering
\includegraphics[width=9cm ]{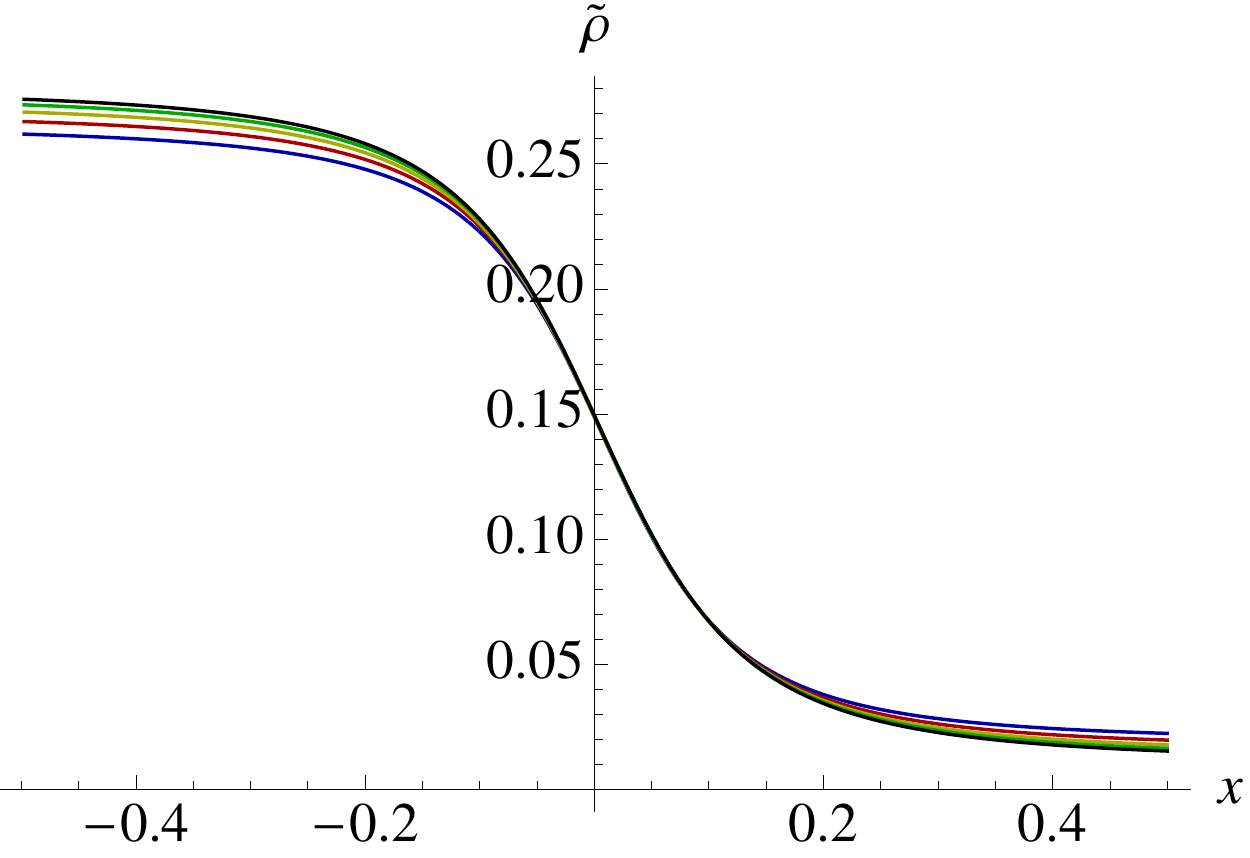}
\includegraphics[width=9cm ]{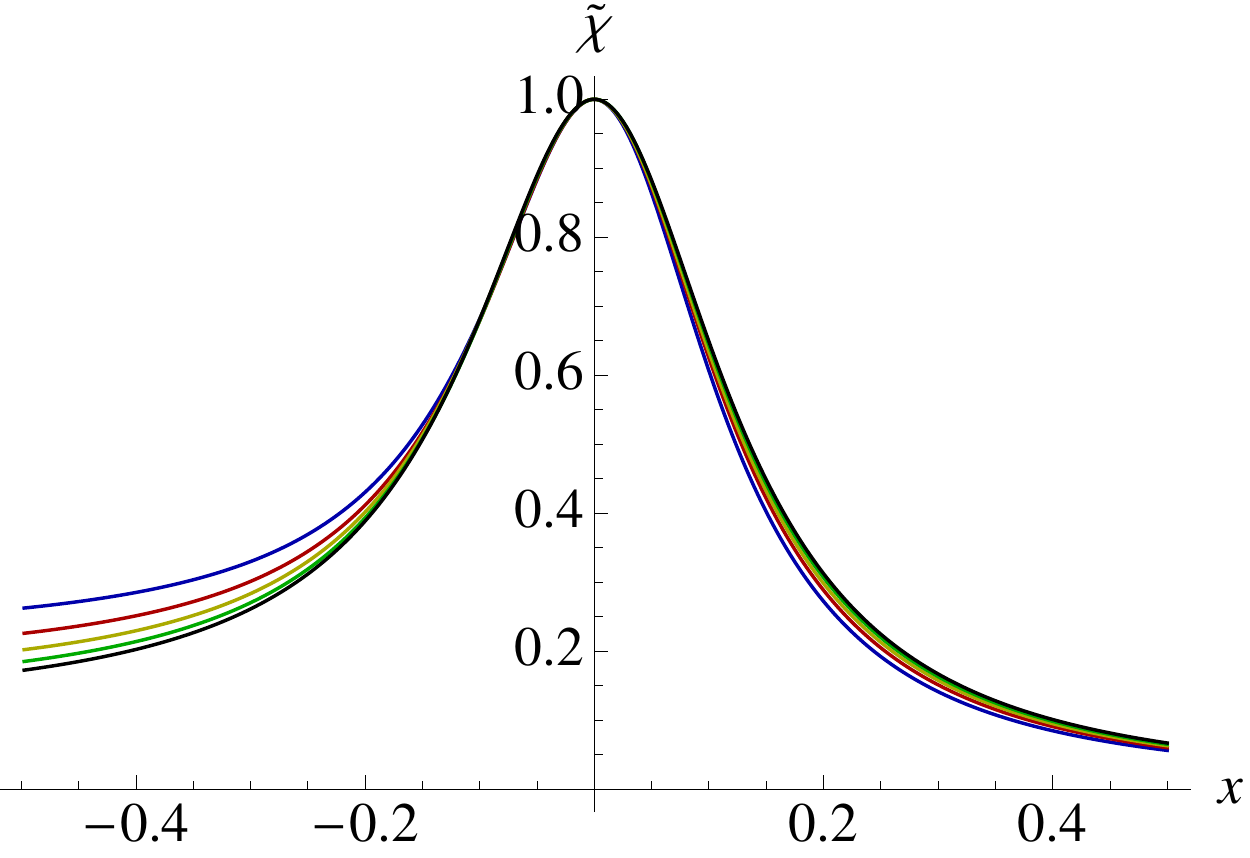}
\caption{\label{Scalingfunctions}
The scaling functions $\tilde \rho(x)= \rho(x \kappa^{-1} + s_\rc) $ (top) and $\tilde \chi(x)= \chi(x\kappa^{-1} + s_\rc)/\chi(s_\rc)$ (bottom) for $c=0.3$ and for various~$L$. The \textcolor{darkblue}{blue}, \textcolor{darkred}{red}, \textcolor{darkgreen}{green}, \textcolor{darkyellow}{yellow}, and black lines correspond to $L=60,70,80,90$, and $L=100$, respectively.
}
\end{figure}

Next, we seek for the scaling functions of $ \rho(s)$ and $\chi(s)$ in the coexistence region. For this purpose, 
we introduce a scaling variable $x$ as
\begin{equation}
x =\kappa (s-s_\rc).
\label{definitionx}
\end{equation}
Correspondingly, the rescaled expectation value and variance are defined as 
\begin{equation}
\tilde \rho(x) = \rho(x \kappa^{-1} + s_\rc)
\label{RescaledMeanOccupation}
\end{equation}
and 
\begin{equation}
\tilde \chi (x) = \chi(x\kappa^{-1} + s_\rc)/\chi(s_\rc),
\label{RescaleOccupationVariance}
\end{equation}
respectively. These functions are plotted for large values of $L$ in Fig.~\ref{Scalingfunctions}, thus illustrating their expected collapse as $L$ becomes large.

\subsection{A variational formula in the infinite system-size limit}
\label{subsec:var-inf-size}
In the previous section, we found numerically that the scaling functions are well-defined with an exponential rescaling with the system size. Hereafter, we discuss how to derive those scaling properties analytically. With this purpose in mind, we consider the variational principle~(\ref{variationalprinciple}). 
We introduce a variational function $\tilde \Phi_{\rm L}(n)$ as
\begin{equation}
\tilde \Phi_{\rm L}(n)=\Phi^{\tomax}(n) P_{\rm eq}(n)^{-1/2}. 
\label{variationalfunction_PhiL(n)}
\end{equation}
By changing the variational parameter $\Phi^{\tomax}$ to $\tilde \Phi_{\rm L}$, we rewrite~(\ref{variationalprinciple}) as
\begin{equation}
L \psi (s) = \max _{\tilde \Phi_{\rm L}>0} \sum_{n^{\prime}}\widetilde P(n^{\prime})  \sum_{n} \tilde \Phi_{\rm L}(n) \evol_{n,n^{\prime}} \tilde \Phi_{\rm L}(n^{\prime})^{-1},
\label{variationalprinciple2}
\end{equation}
where the distribution function $\widetilde P(n)$ is defined as
\begin{equation}
\widetilde P(n) = \frac{ P_{\rm eq}(n)\tilde \Phi_{\rm L}(n)^2}{\sum_{\tilde n}P_{\rm eq}(\tilde n)\tilde \Phi_{\rm L}(\tilde n)^2}.
\label{Peq_prime}
\end{equation}
We note that since, in~(\ref{variationalprinciple}), the optimal $\tilde \Phi_{\rm L}$ corresponds to the largest eigenvector~$\Phi_{\rm L}$, the optimal distribution function $\widetilde P$ is equal to the modified equilibrium distribution function~(\ref{distributionfunction_s}). 
Extensions of such variational principle are valid not only for equilibrium dynamics but also for nonequilibrium dynamics that violate the detailed balance condition. We refer the reader to Ref.~\cite{nemoto_thermodynamic_2011} for details.
In this section, before considering the finite-size scaling properties, we start by considering the $L\rightarrow \infty$ limit by assuming a large deviation principle in the optimal modified system holds.

Let us now introduce, similarly to~\eqref{HamiDiff_def}, a variational function $\Delta \widetilde F$ as
\begin{equation}
\Delta \widetilde F(n) = -2 \log \tilde \Phi_{\rm L}(n),
\label{VariationalFreeEnergyDifference}
\end{equation}
and change the variational parameter $\tilde \Phi_{\rm L}$ to $\Delta \widetilde F$. We note that the optimal $\Delta \widetilde F$ is equal to the free energy difference $\Delta F_{s}$ defined in~(\ref{HamiDiff_def}).
The extremalization principle~(\ref{variationalprinciple2}) can be rewritten as 
\begin{equation}
\psi (s)  = \frac{1}{L}\max _{\Delta \widetilde F} \sum_{n} \widetilde P(n) \left [ \tilde r(n) -r(n) \right ],
\label{variationalprinciple3}
\end{equation}
where $\tilde r(n)$ is defined as
\begin{eqnarray}
\tilde r(n) &= &n c\left ( 1-\frac{n}{L} \right ) \ee^{-\frac 12 [\Delta \widetilde F(n+1)-\Delta \widetilde F(n)] -s} 
\nonumber \\ &&
 \ + n(1-c)\left ( \frac{n}{L} - \frac{1}{L} \right ) \ee^{-\frac 12 [\Delta \widetilde F(n-1)-\Delta \widetilde F(n)] -s}.
\label{rprimedef}
\end{eqnarray}
Now, we assume that the optimal modified distribution $\widetilde P$ scales, in the large~$L$ asymptotics, by following the large deviation form
\begin{equation}
\widetilde P(n) \sim \ee^{-L \left [f_{\rm e}(n/L)+\tilde f(n/L)\right ]},
\label{VariationalFunctionForDensityFreeEnergyDifference}
\end{equation}
where $\tilde f(\rho)$ is an unknown function we determine below. This indicates that the variational function $\Delta \widetilde F(n)$ scales as $L \tilde f(\rho)$ with $\rho=n/L$, as $L\to\infty$.
%Furthermore, we also assume that the large deviation function reaches its minimum at only one point. 
From these assumptions, (\ref{rprimedef}) is rewritten as
\begin{equation}
\frac{\tilde r(L\rho)}{L}=\ee^{-s}\left [\rho c\left ( 1-\rho \right ) \ee^{-\frac 12 {\tilde f}'\!(\rho)}
  + \rho^2 (1-c) \ee^{\frac 12  {\tilde f}'\!(\rho)} \right ]+ O(1/L)
\label{eq:r0largeL}
\end{equation}
where ${\tilde f}'\!(\rho)=\frac{\partial \tilde f(\rho)} {\partial \rho}$.
By substituting this into~(\ref{variationalprinciple3}) and evaluating the summation within the saddle point approximation, we obtain, up to terms in $O( 1/L)$
\begin{eqnarray}
&\hspace*{-2cm} \psi (s) \nonumber\\
& \hspace*{-2cm} = \max _{\tilde f>0} 
\frac
{\displaystyle\int \! d\rho\, \Big\{
\ee^{-s}\Big [ \rho c\big ( 1-\rho \big ) \ee^{-\frac 12 {\tilde f}'\!(\rho)  } + \rho^2 (1-c)\ee^{\frac 12 {\tilde f}'\!(\rho)}\Big ] 
-r(L \rho)/L
\Big\} \ee^{-L[f_{\rm e}(\rho)+\tilde f(\rho)]}
 }
{\displaystyle\int \! d\rho\, \ee^{-L[f_{\rm e}(\rho)+\tilde f(\rho)]} }
%+ O(L^{-1}),
\label{variationalprinciple40}
\end{eqnarray}
Here $f_{\rm e}(\rho)$ is the large deviation function~\eqref{eq:refIequilib}.
We now evaluate the integrals over $\rho$ through the saddle-point method.
From~\eqref{eq:r0largeL}, we read that the exponential dependence in $L$ of the numerator and the denominator of~\eqref{variationalprinciple40} is the same. Assuming that the optimal function $f_{\rm e}(\rho)+\tilde f(\rho)$ reaches its minimum at a unique point $\rho^{\tomax}$, we have that the numerator and the denominator of~\eqref{variationalprinciple40} are both dominated by $\rho=\rho^{\tomax}$. With the condition
\begin{equation}
{\tilde f}'\!(\rho^{\tomax}) +{f_{\rm e}}'(\rho^{\tomax})=0,
\label{Saddlecondition}
\end{equation}
this leads to the variational principle
\begin{equation}
\hspace*{-2cm}
\psi (s)  = \max _{0\leq\rho^{\tomax}\leq 1} 
\Big\{
\ee^{-s}\Big [ \rho^{\tomax} c\big ( 1-\rho^{\tomax} \big ) \ee^{-\frac 12 {\tilde f}'\!(\rho^{\tomax})  } + (\rho^{\tomax})^2 (1-c)\ee^{\frac 12 {\tilde f}'\!(\rho^{\tomax})}\Big ] 
-r(L \rho^{\tomax})/L
\Big\}
\label{variationalprinciple4}
\end{equation}
Finally, by substituting~(\ref{Saddlecondition}) into~(\ref{variationalprinciple4}) and using the definition~\eqref{eq:refIequilib} of $f_{\rm e}(\rho)$, we arrive at
\begin{equation}
\hspace*{-2cm}
\psi (s)  = \max _{0\leq\rho\leq 1} 
\left \{
2 \ee^{-s} \sqrt{\rho^3(1-\rho) c(1-c)} -\left [ \rho c (1-\rho) + (1-c)\rho^2\right ]
\right \} + O(1/L)
\label{variationalprinciple5}
\end{equation}
Such variational formula describes the large-size behavior of large deviation of a wide class of stochastic models on the complete graph~\cite{lecomte_thermodynamic_2007,garrahanjacklecomtepitardvanduijvendijkvanwijland,garrahan_first-order_2009}
or quantum annealing models~\cite{1742-5468-2012-06-P06007}.
% The formula was obtained by assuming the large deviation property and the existence of only one saddle in the optimal modified system.
In Fig.~\ref{DynamicalFreeEnergy}, the obtained $\lim_{L\to\infty} \psi(s)$, whose transitions occurs at $s_\rc^\infty=0$, is compared to finite-size results obtained by numerical diagonalization.

We now discuss how to extend the discussion presented in this subsection so as to obtain the finite-size scaling.
A blunt perturbation in $1/L$ around~(\ref{variationalprinciple5}) cannot be performed directly, since for instance it cannot describe how the $O(L^0)$ discontinuity of $K(s)$ at $s_c^\infty=0$ is rounded at large but finite~$L$.
More generically, the free energy difference $\Delta F_{s}$ defined in~\eqref{HamiDiff_def}, which allows one to recover $\psi(s)$, is exactly 0 at $s=0$. 
One could thus expect that $\Delta F_s/L=O(1/L)$ at the transition point, using that $s_\rc=O( 1/L)$ (Section~\ref{roughs_c}).
However, because of the non-analyticity of $\Delta F_s/L$ with respect to $s$ around the transition point $s_\rc$,
one has in fact $\Delta F_s/L=O(1)$, as discussed in Section~\ref{AnalyticalHamiDiff}.
This explains the breakdown of a $ 1/L$ expansion of~(\ref{variationalprinciple5}). In the next section, we propose a new method to overcome those difficulties without relying on a perturbative approach.

%  A perturbation in $\frac 1L$ around~(\ref{variationalprinciple5}) will work. However, it will not work because of the following reason. When $s=0$, we know that the optimal modified system is equivalent to the original system so that the optimal modified Hamiltonian $\Delta H_{s}/L$ (large deviation function) is exactly 0. Furthermore, we know that $s_\rc$ is $O(1/L)$ as shown in Section~\ref{roughs_c}. We thus expect that $\Delta H_s/L$ at the transition point
% is also $O(1/L)$ by considering the taylor expansion of $\Delta H_s/L$ around $s=0$.
% It is not true, however, as shown in Section~\ref{AnalyticalHamiDiff}, which means $\Delta H_s/L=O(1)$. This is because of the non-analyticity of $\Delta H_s/L$ with respect to $s$ around the transition point $s_\rc$.
% Since $\Delta H_{s}/L$ is $O(1)$, the difference between optimal modified system at the transition point and the original system is also $O(1)$. Thus, the perturbation theory around~(\ref{variationalprinciple5}) will not work to describe the scaling behaviors at the transition point. In the next section, we propose a new method to overcome those difficulties without relying on perturbative approach.

\subsection{Analytical expressions for the scaling functions}
\label{analyticalscalingfunctions}
Let us first consider the coexistence point $s=s_\rc$. 
We use the finite-size correction of the free energy obtained in Section~\ref{AnalyticalHamiDiff}. 
We first define the finite-size free energy difference for each region as
\begin{equation}
f_{\rm i}(\rho)=\Delta f_{s_{\rc}}(\rho) +\frac{1}{L} \Delta f_{\rm s_c}^{(1)}(\rho)
\label{f_iDefinition}
\end{equation}
for $\rho \leq \rho_{\rm \rc}^L$
\begin{equation}
f_{\rm a}(\rho)=\Delta f_{s_{\rc}}(\rho) +\frac{1}{L} \Delta f_{\rm s_c}^{(1)}(\rho)
\label{f_aDefinition}
\end{equation}
for $\rho > \rho_{\rm \rc}^L$. Then, the corresponding distribution function is 
\begin{equation}
P^{s_\rc}(n)=
P_{\rm i}(n)\,\mathbf{1}_{n\le n_\rc^L}
+
P_{\rm a}(n)\,\mathbf{1}_{n>n_\rc^L},
\end{equation}
where $P_{\rm i}(n)$ and $P_{\rm a}(n)$
are defined as
\begin{equation}
P_{\rm i}(n)=\frac{1}{Z_{\rm i} + Z_{\rm a}} P_{\rm eq}(n) \ee^{-L f_{\rm i}(n/ L)}
\label{P_iDefinition}
\end{equation}
\begin{equation}
P_{\rm a}(n)=\frac{1}{Z_{\rm i} + Z_{\rm a}} P_{\rm eq}(n) \ee^{-L f_{\rm a}(n/L)}
\label{P_aDefinition}
\end{equation}
with the normalization constant
\begin{equation}
Z_{{\rm i}} =\sum_{n\leq n_\rc} P_{\rm eq}(n) \ee^{-L f_{\rm i}(n/ L)},
\end{equation}
\begin{equation}
Z_{{\rm a}} =\sum_{n > n_\rc} P_{\rm eq}(n) \ee^{-L f_{\rm a}(n/ L)}.
\end{equation}
Here, we note $Z_{\rm i}= Z_{\rm a}$ from the definition~\eqref{condition_for_rhoc} of $\rho_\rc^L$. In this system, there naturally arise two regions, $n\leq n_\rc$ and $n>n_\rc$, which represent inactive phase and active phase, respectively.
As seen below, this separation plays an important role in the exponentially fast scaling of Section~\ref{Scalingfunction}.

By using the result at $s=s_\rc$, we consider the region around $s_\rc$ ($s\sim s_\rc$).
Here, we propose an Ansatz. That is, the distribution function around $s_\rc$ is written as
\begin{equation}
P^{s}(n)=[1+a^{*}(s)] \delta_{n\leq n_\rc}P_{\rm i}(n)
+[1-a^{*}(s)] \delta_{n>n_\rc}P_{\rm a}(n),
\label{Ansatz1}
\end{equation}
where $a^{*}(s)$ is a `mixing function' to determine.
This Ansatz is inspired by the mathematically-proved description of static and classical equilibrium first-order transitions in the coexistence region as described in~\cite{borgskotecky,PhysRevLett.68.1734} based on a Gibbs distribution picture. We should mention that in contrast to~\cite{borgskotecky,PhysRevLett.68.1734}, our full distribution $P^{s}(n)$ does not describe a superposition between two elementary distributions, but a separation of the space of occupation number~$n$ into two regions presenting distributions of different nature. Also, we stress that the main difference between our method and their method appears in the mixing function $a^*(s)$, which is be determined by a variational principle as shown below. The dynamical thermodynamic formalism has a variational principle which determines the stationary state, in contrast to the equilibrium thermodynamics. The mixing function $a^*(s)$ reflects this inevitable difference between these two thermodynamics.

To our knowledge, the Ansatz~\eqref{Ansatz1} has not been used up to now in the dynamical approach we follow.
In physical terms, the Ansatz~\eqref{Ansatz1} allows to handle the difficulty at the core of the extension of the infinite-size method presented in Section~\ref{subsec:var-inf-size}. In the infinite-size limit, the variational maximum~\eqref{variationalprinciple2} is dominated by the most probable value of the occupation density $\rho$, yielding~\eqref{variationalprinciple5}. In finite-size this image however breaks down in the vicinity of $s=s_\rc$, where the active and inactive densities $\rho\simeq 0$ and $\rho\simeq c$ are in competition, and finer details than the mere maximum of the distribution of~$\rho$ have to be taken into account at large but finite $L$.
The Ansatz~\eqref{Ansatz1} describes the continuous transition, as $s$ increases, from the active to the inactive \textit{distribution} of density.

The distribution function~\eqref{Ansatz1} satisfies the normalization condition because $Z_{\rm i} = Z_{\rm a}$. From this expression and the relation~(\ref{distributionfunction_s}), we obtain $\Phi_{\rm L}(n)$ (or the 
free energy difference $\Delta F_{s}(n)$) as
\begin{equation}
\Phi_{\rm L}(n) \propto \delta_{n\leq n_\rc} \sqrt{1+ a^{*}(s)} \ee^{-L f_{\rm i}(n/L)/2} + \delta_{n>n_\rc} \sqrt{1- a^{*}(s)}\ee^{-L f_{\rm a}(n/L)/2},
\label{Ansatz2}
\end{equation}
%or
%\begin{equation}
%\Delta H_{\rm s}(n)=\delta_{n\leq n_{c}}\left [ L h_{\rm i}(n/L) -\log(1+a^{*}(s)) \right ]
%+\delta_{n> n_{c}}\left [ L h_{\rm a}(n/L) -\log(1-a^{*}(s)) \right ] +\rm const.
%\end{equation}
and the expectation values $\rho(s)$, $\chi(s)$ around $s_\rc$ as
\begin{equation}
\rho(s)= \frac{\left \langle \rho \right \rangle_{\rm i}}{2}\left [ 1+a^*(s) \right ] 
+ \frac{ \left \langle \rho \right \rangle_{\rm a}}{2}\left [ 1-a^*(s) \right ],
\label{Expectation_variance1}
\end{equation}
\begin{equation}
\chi(s)=L\left \{
 \frac{\left \langle \rho ^2  \right \rangle_{\rm i}}{2}\left [ 1+a^*(s) \right ]+ \frac{\left \langle \rho^2 \right \rangle_{\rm a}}{2} \left [ 1-a^*(s) \right ]- \rho(s)^2
\right \},
\label{Expectation_variance2}
\end{equation}
where $\left \langle \ \right \rangle_{\rm i}$ and $\left \langle \ \right \rangle_{\rm a}$ are 
the expectation values in the active and the inactive phases defined as
$\left \langle g \right \rangle_{\rm i}=2\sum_{n \leq n_\rc}P_{\rm i}(n)g(n)$ and
$\left \langle g \right \rangle_{\rm a}=2\sum_{n>n_\rc}P_{\rm a}(n)g(n)$ respectively.
When we know $a^{*}(s)$, we thus can calculate every scaling property around $s_\rc$.

For determining $a^{*}(s)$, we use the variational principle
(\ref{variationalprinciple3}).
We consider the variational function $\Psi(a)$ defined as
\begin{equation}
\Psi(a)=\frac{1}{L}\sum_{n}P^{s}(n)\left [ \tilde r(n)-r(n)  \right ],
\label{VariationalfunctionPsi}
\end{equation}
with
\begin{equation}
\tilde r(n)=n c \left ( 1 - \frac{n}{L} \right ) \frac{\Phi_{\rm L}(n+1)}{\Phi_{\rm L}(n)}\ee^{-s} +n(1-c) \left ( \frac{n}{L} 
-\frac{1}{L} \right ) \frac{\Phi_{\rm L}(n-1)}{\Phi_{\rm L}(n)}\ee^{-s},
\end{equation}
where $P^{s}$ and $\Phi_{\rm L}$ in these expressions are~(\ref{Ansatz1}) and~(\ref{Ansatz2}) with the replacement of the mixing function $a^{*}(s)$ by a mixing parameter $a$. We determine $a^{*}(s)$ from the
condition
\begin{equation}
\partial \Psi(a)/\partial a|_{a=a^*(s)}=0.
\label{eq:variational_a}
\end{equation}
For evaluating the variational function, we divide the region of the summation in~(\ref{VariationalfunctionPsi}) into three parts, (i) $n<n_\rc$, (ii) $n>n_\rc+1$, and (iii)
$n=n_\rc$, $n_\rc+1$. For this, we define functions
\begin{equation}
\Psi_{<}(a)=(1/L)\sum_{n<n_\rc} P^{s}(n)\left [ \tilde r(n) -r(n)\right ],
\end{equation}
\begin{equation}
\Psi_{>}(a)=(1/L)\sum_{n>n_\rc} P^{s}(n)\left [ \tilde r(n) -r(n)\right ],
\end{equation}
\begin{equation}
\Psi_{=}(a)=(1/L) \sum_{n=n_\rc,n_\rc+1} P^{s}(n)\left [ \tilde r(n) -r(n)\right ].
\end{equation}
For (i) and (ii), one can easily find that the dependence in the mixing parameter $a$ is linear because $\tilde r(n)$ 
is an independent function of $a$.
Indeed, we obtain
\begin{equation}
\Psi_{<}(a)=(1+a)\Omega_{<},
\end{equation}
\begin{equation}
\Psi_{>}(a)=(1-a)\Omega_{>},
\end{equation}
where $\Omega_{<}$ and $\Omega_{>}$ are constants defined as
\begin{equation}
\Omega_{<}=\frac{1}{2L}\left \langle \tilde r_{\rm i}\ee^{-s} - r  \right \rangle_{\rm i},
\label{Omegasmalldef}
\end{equation}
\begin{equation}
\Omega_{>}=\frac{1}{2L}\left \langle \tilde r_{\rm a}\ee^{-s} - r  \right \rangle_{\rm a},
\label{Omegalargedef}
\end{equation}
with the definition of $\tilde r_{\rm i,a}(n)$ as
\begin{eqnarray}
\tilde r_{\rm i,a}(n)= & n c \left ( 1 - \frac{n}{L} \right ) \frac{\ee^{-L f_{\rm i,a}((n+1)/L)/2} }{\ee^{-L f_{\rm i,a}(n/L)/2} } \nonumber \\
& +n(1-c) \left ( \frac{n}{L} 
-\frac{1}{L} \right ) \frac{\ee^{-L f_{\rm i,a}((n-1)/L)/2} }{\ee^{-L f_{\rm i,a}(n/L)/2} }.
\end{eqnarray}
Different from these parts, however, $\Psi_{=}(a)$ does not depend  linearly on the mixing parameter 
$a$. Because of this fact, we have to consider this part in spite of it being  exponentially small compared with $\Psi_{>}(a)$ and $\Psi_{<}(a)$. $\Psi_{=}(a)$ is
\begin{eqnarray}
\Psi_{=}(a) = & \frac{n_\rc}{L} c(1-\frac{n_\rc}{L}) \frac{\Phi_{\rm L}(n_\rc+1)}{\Phi_{\rm L}(n_\rc)}P^{s}(n_\rc) \ee^{-s} \nonumber \\
& + \frac{(n_{c}+1)}{L}(1-c)\frac{n_\rc}{L}\frac{\Phi_{\rm L}(n_\rc)}{\Phi_{\rm L}(n_\rc+1)}P^{s}(n_\rc+1) \ee^{-s} + \cdots ,
\label{psi=}
\end{eqnarray}
where $\cdots$ stands for the terms that are proportional to $a$. 
The non-linear dependence in the mixing parameter~$a$ is crucial (without it, the variational principle~\eqref{eq:variational_a} would be of no use) and arises from the finite-size correction to the free energy difference obtained at the end of Section~\ref{AnalyticalHamiDiff}.

By using $P^{s}(n+1)\Phi_{\rm L}(n)/\Phi_{\rm L}(n+1)= P^{s}(n) \Phi_{\rm L}(n+1) P_{\rm eq}(n+1)/(\Phi_{\rm L}(n)P_{\rm eq}(n))$, we find that the first term and the second term in the right hand side of~(\ref{psi=}) are equal. We thus have
\begin{equation}
\Psi_{=}(a) = 2 \sqrt{1-a^2} \frac{n_\rc}{L} c(1-\frac{n_\rc}{L}) P_{\rm i}(n_\rc) \frac{\ee^{-L f_{\rm a}((n_\rc+1)/L)/2} }{\ee^{-L f_{\rm i}(n_\rc/L)/2} }\ee^{-s}+\cdots.
\end{equation}
By defining constants 
\begin{equation}
\Omega_{=}=2  \frac{n_\rc}{L} c(1-\frac{n_\rc}{L}) P_{\rm i}(n_\rc) \frac{\ee^{-L f_{\rm a}((n_\rc+1)/L)/2} }{\ee^{-L f_{\rm i}(n_\rc/L)/2} }\ee^{-s},
\end{equation}
we obtain an expression of $\Psi(a)$ that represents the $a$ dependence. That is,
\begin{equation}
\Psi(a)=\Omega_{<}+\Omega_{>} + a (\Omega_{<}-\Omega_{>}) + \sqrt{1-a^2}\Omega_{=}.
\end{equation}
Thus, by maximizing $\Psi(a)$ with respect to $a$, we arrive at the expression of $a^*(s)$
as
\begin{equation}
a^*(s)=\frac{A}{\sqrt{1+A^2}},
\label{a*(s)}
\end{equation}
where $A$ is 
\begin{equation}
A=(\Omega_{<}-\Omega_{>})/\Omega_{=}.
\label{DefA}
\end{equation}

Now, by using~(\ref{a*(s)}) and~(\ref{DefA}), we discuss the scaling properties.
First, we focus on the scaling ratio $\kappa$ defined in~(\ref{ScalingRationDef}).
We know that $a^*(s_\rc)=0$ by definition. From
(\ref{Omegasmalldef}), (\ref{Omegalargedef}), (\ref{a*(s)}) and~(\ref{DefA}),
 we thus obtain an equation for $s_\rc$ as
\begin{equation}
\frac{1}{L}\left \langle \tilde r_{\rm i}\ee^{-s_\rc} - r  \right \rangle_{\rm i}
=\frac{1}{L}\left \langle \tilde r_{\rm a}\ee^{-s_\rc} - r  \right \rangle_{\rm a}.
\label{Conditionfors_c}
\end{equation}
Then, we expand $A$ around $s=s_\rc$ by using this condition.
By denoting $\Omega_{=}|_{s=s_\rc}$ by $\Omega_{=}^{\rm c}$, we obtain
\begin{equation}
A=- \frac{s-s_\rc}{\Omega_{=}^{\rm c}}\left [ \left \langle \frac{r}{2L} \right \rangle_{\rm i} - \left \langle \frac{r}{2L} \right \rangle_{\rm a} \right ]  + O((s-s_\rc)^2),
\label{ExpressionA}
\end{equation}
from which, with~(\ref{Expectation_variance1})
and~(\ref{a*(s)}), we find the scaling ratio $\kappa=-\partial \rho(s)/\partial s|_{s=s_\rc}$ as
\begin{equation}
\kappa = \frac{1}{\Omega_{=}^{\rm c}}\left [ \left \langle
\frac{\rho}{2}
 \right \rangle_{\rm i} -\left \langle
\frac{\rho}{2}
 \right \rangle_{\rm a} \right ]\left [ \left \langle
\frac{r}{2L}
 \right \rangle_{\rm i} -\left \langle
\frac{r}{2L}
 \right \rangle_{\rm a} \right ].
 \label{Expressionkappa}
\end{equation}
Here, we note that the~$L$ dependence in $\kappa$ mainly comes from $P_{\rm i}(n_\rc)$ in $\Omega_{=}^{\rm c}$ because each of the other terms converges to each corresponding value in the $L \rightarrow \infty$ limit. 
That is, when~$L$ is large, $\log \kappa \rightarrow - \log P_{\rm i}(n_\rc) + \rm const.$ By using the large deviation property of $P_{\rm i}(n_\rc)$, we thus arrive at
\begin{equation}
\frac{1}{L}\log \kappa \longrightarrow    f_{s_{\rm c}}(\rho_\rc^{\infty})= -\frac{1}{2}\log (1-c)
\qquad\text{as $\quad L\to\infty$}
.
\label{kapparesult}
\end{equation}
We note that the slope of the straight lines in Fig.~\ref{ScalingRatio} is the height of the large deviation function from bottom to the connecting point ($\rho=\rho_\rc^{\infty}$). This reminds us the instantonic approach used in~\cite{0295-5075-89-4-40004}. Along a similar vein, Bapst and Semerjian derived a formula determining the exponentially small gap in the quantum ferromagnet~\cite{1742-5468-2012-06-P06007}. In the next section, by using our approach, we will re-derive this formula.

\begin{figure}[h]
\centering
\includegraphics[width=9cm ]{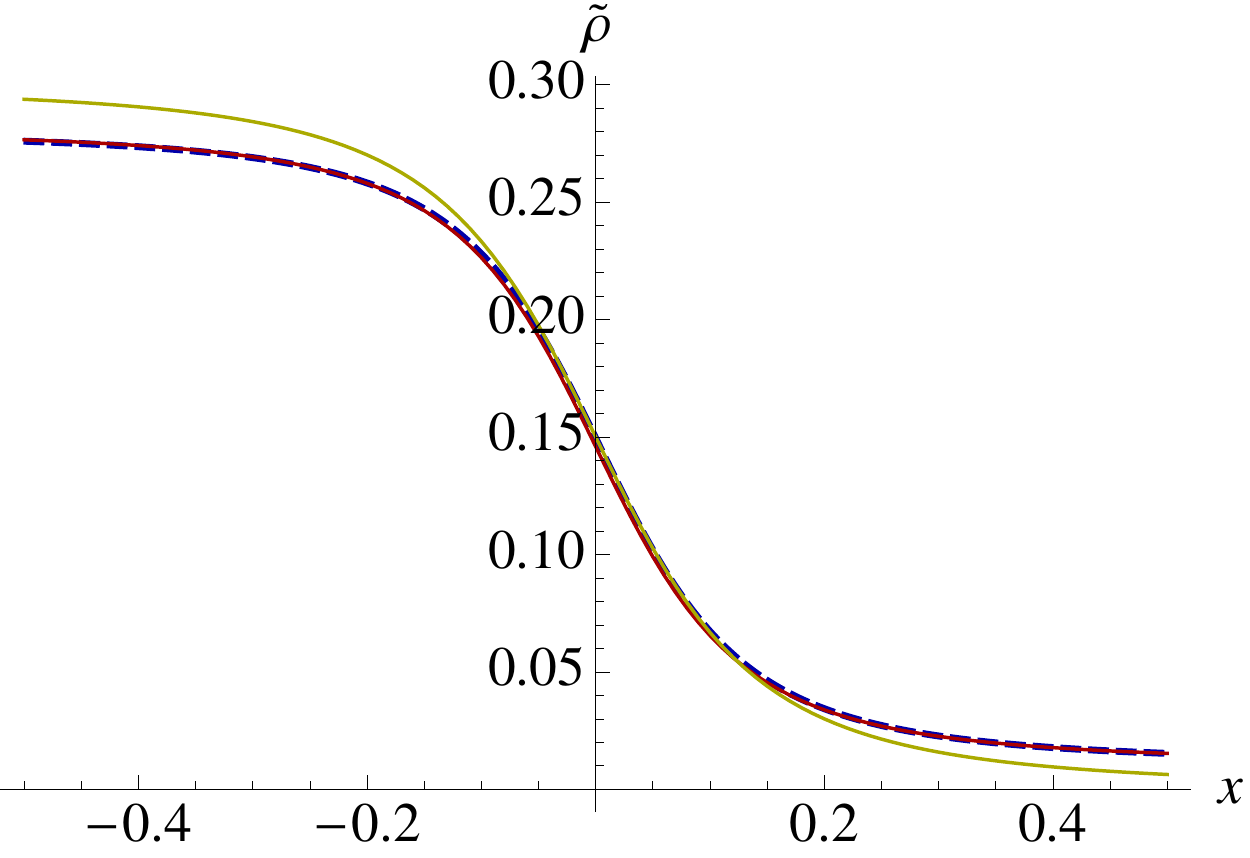}
\includegraphics[width=9cm ]{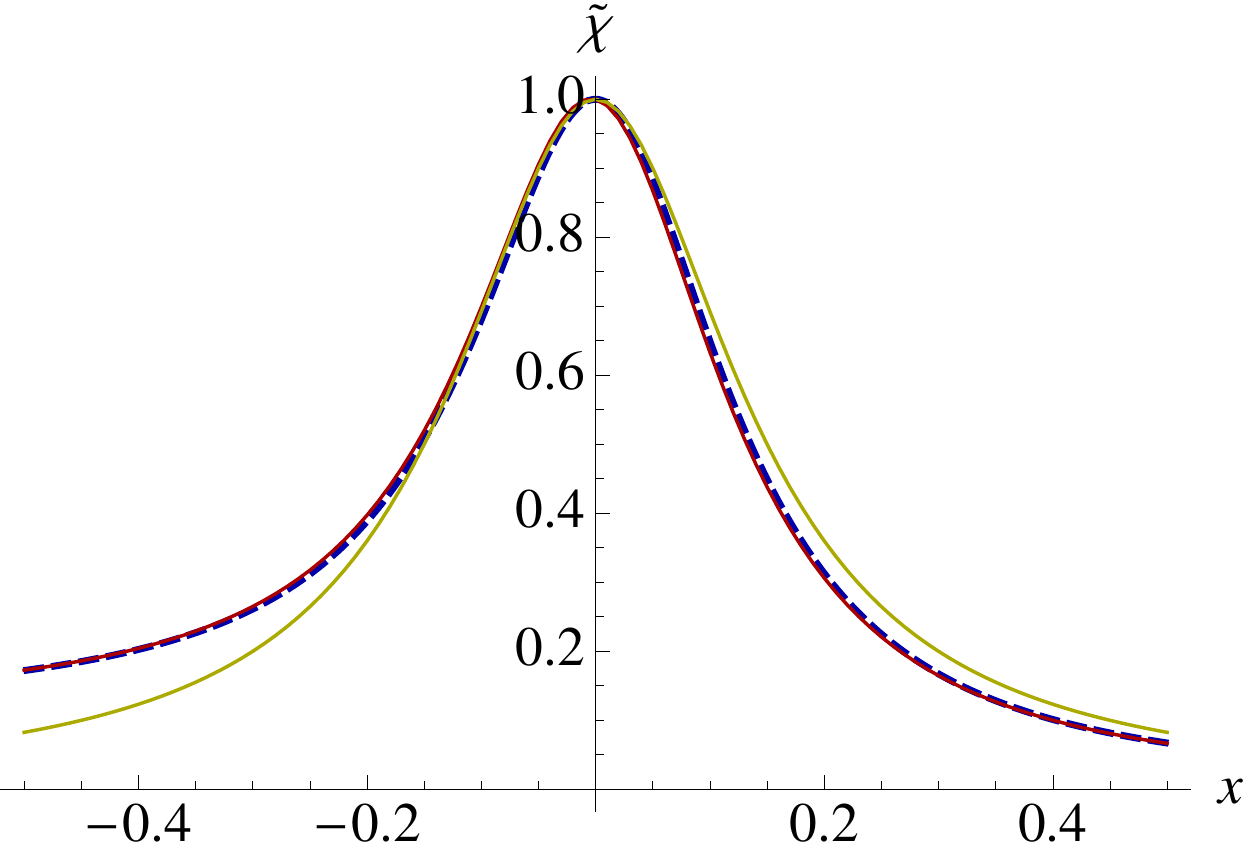}
\caption{\label{ScalingfunctionsAnaly}Analytical results for the scaling functions $\tilde \rho(x)= \rho(x \kappa^{-1} + s_\rc) $ (up) and $\chi(x)= \chi(x\kappa^{-1} + s_\rc)/\chi(s_\rc)$ (down) for $c=0.3$ for $L=100$. The solid {\textcolor{darkred}{red}} lines are the analytical results~(\ref{analyscalingfunction1})
and~(\ref{analyscalingfunction2}). The dashed {\textcolor{darkblue}{blue}} lines are the numerical results, which are the same as in Fig.~\ref{Scalingfunctions}.
We also plot the infinite-size scaling functions~(\ref{analyscalingfunction1_infinite}) and~(\ref{analyscalingfunction2_infinite}) as the {\textcolor{darkyellow}{yellow}} lines, which illustrate the importance of large but finite-size contributions.
}
\end{figure}

Next, we obtain the expression of the scaling function.
By combining the definition of $x$ in~(\ref{definitionx}) with~(\ref{ExpressionA}) and~(\ref{Expressionkappa}), we know
\begin{equation}
A=\frac{2x}{
\left \langle
\rho
 \right \rangle_{\rm a} -\left \langle
\rho
 \right \rangle_{\rm i} } + O(\kappa^{-1}).
 \label{A_as_Function_Of_x}
\end{equation}
From this with~(\ref{Expectation_variance1}),~(\ref{Expectation_variance2}) and~(\ref{a*(s)}),
we find 
the analytical expression of $\tilde \rho(x)$ and $\tilde \chi(x)$ as
\begin{equation}
\tilde \rho(x)=\frac{1}{2} \left [
\left \langle \rho\right \rangle_{\rm i}
+\left \langle \rho\right \rangle_{\rm a}
- \frac{2x}{\sqrt{1+4x^2\left [\left \langle \rho\right \rangle_{\rm i}
-\left \langle \rho\right \rangle_{\rm a} \right ]^{-2} }}
\right ],
\label{analyscalingfunction1}
\end{equation}
\begin{eqnarray}
\tilde \chi(x) & =  \frac{1}
{\left \langle \rho^2 \right \rangle_{\rm i}+\left \langle \rho^2 \right \rangle_{\rm a} - \left [ \left \langle \rho \right \rangle_{\rm i}+\left \langle \rho \right \rangle_{\rm a}\right ]^2 /2} \nonumber \\
& \times \left [
\left \langle \rho^2\right \rangle_{\rm i}
+\left \langle \rho^2\right \rangle_{\rm a}
-   \frac{2x \left [\left \langle \rho^2\right \rangle_{\rm i}
-\left \langle \rho^2\right \rangle_{\rm a} \right ]
 \left [\left \langle  \rho \right \rangle_{\rm i}
-\left \langle  \rho \right \rangle_{\rm a} \right ]^{-1}}
{\sqrt{ 1+4x^2\left [\left \langle  \rho \right \rangle_{\rm i}
-\left \langle \rho \right \rangle_{\rm a} \right ]^{-2} }} - 2 \tilde \rho(x)^2 \right ],
\label{analyscalingfunction2}
\end{eqnarray}
where we omit the exponentially small deviation $O(\kappa^{-1})$.
The only parameters appearing in these expressions
are the expectation value and the variance of $\rho$ in each of the active and inactive phases.
They can still bear a finite-size dependency, which is important for numerical analysis at large but finite~$L$, as illustrated in Fig.~\ref{ScalingfunctionsAnaly}.
Because these parameters converge in the $L\rightarrow \infty$ limit,
each of the expressions~(\ref{analyscalingfunction1}) and~(\ref{analyscalingfunction2}) also converge to a limit function. Indeed, from the modified free energy~(\ref{largedeviation_modifiedhamiltonian1}) and~(\ref{largedeviation_modifiedhamiltonian2}), we obtain $\lim_{L\rightarrow \infty}\left \langle \rho \right \rangle_{\rm i}=0$ and $\lim_{L\rightarrow \infty} \left \langle \rho \right \rangle_{\rm a}=c$. This leads to the following infinite-size scaling functions:
\begin{equation}
\tilde \rho_{\infty}(x)=\lim_{L\rightarrow \infty}\tilde \rho(x)
=\frac{1}{2} \left [
c - \frac{2x}{\sqrt{1+4x^2c^{-2} }}
\right ],
\label{analyscalingfunction1_infinite}
\end{equation}
\begin{equation}
\tilde \chi_{\infty}(x)=\lim_{L\rightarrow \infty}\tilde \chi(x)
= \frac{c^2}{c^2+4 x^2}.
\label{analyscalingfunction2_infinite}
\end{equation}
We stress that for the derivation of~(\ref{kapparesult}), (\ref{analyscalingfunction1}) and~(\ref{analyscalingfunction2}), we haven't used the details of this system, e.g. $r(n)$, $f_{\rm i}(\rho)$, $f_{\rm a}(\rho)$. It indicates that the scaling results~(\ref{kapparesult}), (\ref{analyscalingfunction1}) and~(\ref{analyscalingfunction2}) should
 also hold in other systems. We will see an example of such extension to other systems in part~\ref{sec:MFqf}, for a mean-field quantum ferromagnet.

\begin{figure}[h]
\centering
\includegraphics[width=9cm ]{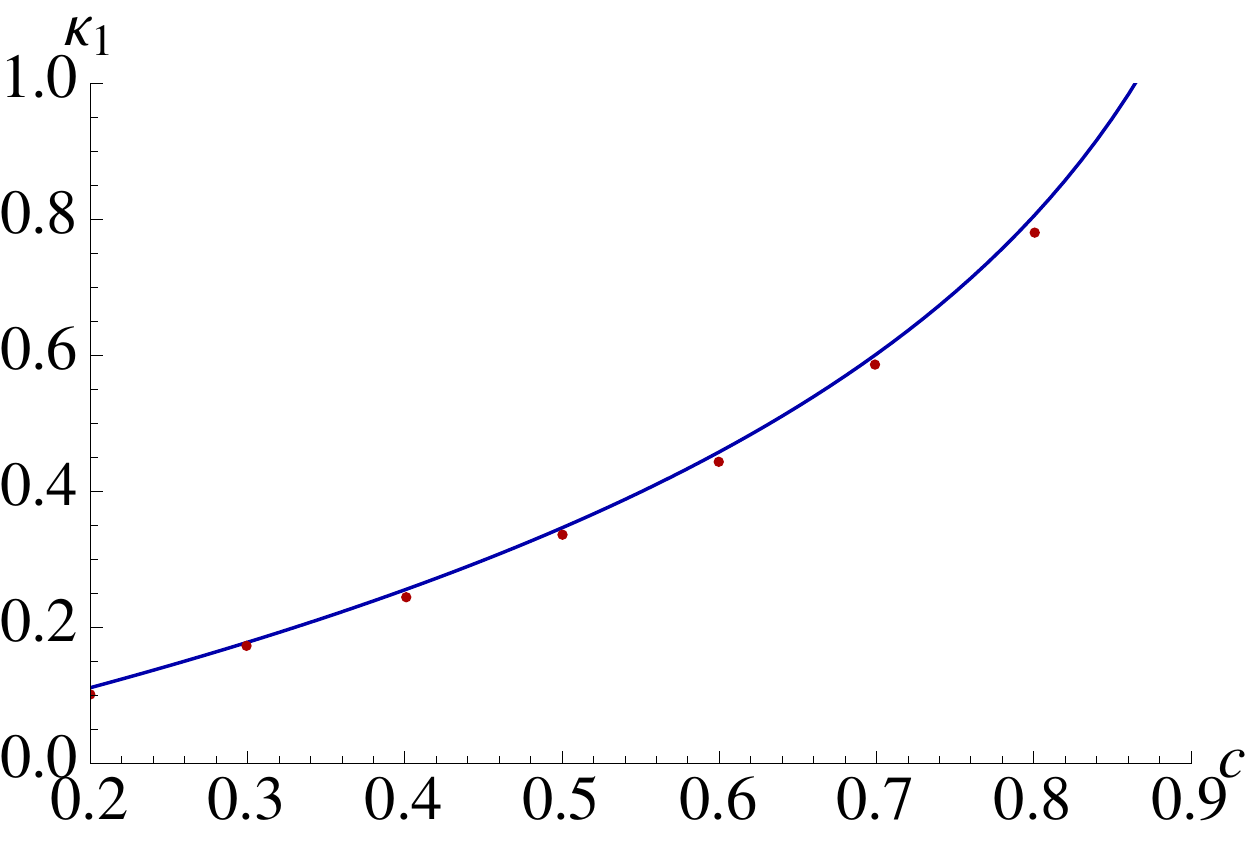}
\caption{\label{FigExponentAnaly}$\kappa_1$ (the exponent in the exponential behavior of $\kappa\sim e^{\kappa_1 L}$). We estimated $\kappa_1$ for $c=0.2,0.3,\dots,0.8$ from Fig.~\ref{ScalingRatio} for finite $L$ and plot those as the red dots. Also, we draw the analytical result $-\frac{1}{2}\log(1-c)$ ({\textcolor{darkblue}{blue line}}) expected in the infinite $L$ limit, see~\eqref{kapparesult}. The small discrepancy corresponds to finite-size effects, see~\eqref{scExpression}. }
\end{figure}

Finally, we numerically check the obtained results~(\ref{kapparesult}), (\ref{analyscalingfunction1}),
and~(\ref{analyscalingfunction2}).
From Fig.~\ref{ScalingRatio}, we estimate the slopes of the straight lines of $\log \kappa$. We denote it by $\kappa_1$.
We plot $\kappa_1$ for $c=0.2,0.3,\dots,0.8$ in Fig.~\ref{FigExponentAnaly} (red dots) and at the same time we also draw $(-1/2)\log(1-c)$ as a function of $c$ (blue line), which is~(\ref{kapparesult}). We can see how they coincide.
Next, 
in Fig.~\ref{ScalingfunctionsAnaly}, we plot the analytical expressions~(\ref{analyscalingfunction1}), (\ref{analyscalingfunction2}) and the corresponding numerical results, as blue dotted lines and red solid lines, respectively. 
The accuracy of the coincidence between the numerical and the analytical results is amazing. 
On the same figure, we also plot the infinite-size scaling functions~(\ref{analyscalingfunction1_infinite}) and~(\ref{analyscalingfunction2_infinite}) in yellow lines. One can see the deviation between the infinite ones and the finite ones. This indicates that larger system sizes are required to observe the convergence to the infinite-size scaling functions. It is worth mentioning that even for the relatively small system sizes, the finite-size scaling formul\ae~(\ref{analyscalingfunction1}) and~(\ref{analyscalingfunction2}) show good agreement with the numerical lines. Because the numerical diagonalization becomes harder as the system size becomes larger, the finite-size scaling functions~(\ref{analyscalingfunction1}) and~(\ref{analyscalingfunction2}) make the check of the validity of our formulation easier.
Furthermore, we also evaluate $s_\rc$ by using~(\ref{Conditionfors_c}). The result is 
\begin{equation}
s_\rc = 1/(2Lc(1-c)) + O(1/L^2),
\label{scExpression}
\end{equation}
which is checked numerically in Fig.~\ref{DynamicalFreeEnergy}. See 
Appendix~\ref{app:sc_expression_derivation} for the details of the evaluation. This tells us that the upper bound in~\eqref{criticlam} is actually saturated.

\subsection{Scaling function of $\partial \psi(s)/\partial s$}
\label{Sec/Scaling function of psi(s)}

So far, we have focused on the scaling property of $\rho(s)$ and $\chi(s)$. In the similar vein, in this section, we will show the scaling function of $\partial \psi(s)/\partial s$ and $\partial^2 \psi(s)/\partial s^2$, which correspond to the expectation value and the susceptibility of the activity in the modified system.

First, the expectation value of the activity in the modified system can be calculated as
\begin{equation}
-\frac{\partial \psi(s)}{\partial s} = \sum _{n} \sum_{n^{\prime}}P^{s}(n) w(n\rightarrow n^{\prime})=\sum_{n}P^{s}(n) r(n),
\end{equation}
where $r(n)$ is the escape rate given as~(\ref{escapedefine}).
Since only $n$ and $n^2$ terms constitute $r(n)$, the expectation value of $\partial \psi(s) / \partial s$ can be expressed by using only $\rho(s)$ and $\chi(s)$.
Indeed, with relations $\rho(s)L = \left \langle n \right \rangle_{\rm eq}^{s}$ and $\left \langle n^2 \right \rangle_{\rm eq}^{s}= L \chi(s) + L^2\rho(s)^2$, we obtain
\begin{equation}
-\frac{\partial \psi(s)}{\partial s} = \rho(s) (Lc+c-1) + \rho(s)^2 L (1-2c) + (1-2c) \chi(s).
\end{equation}
%and 
%\begin{equation}
%\frac{\partial^2 \psi(s)}{\partial s^2} = \frac{\partial \rho(s)}{\partial s} (Lc+c-1) + 2 \rho(s) \frac{\partial \rho(s)}{\partial s} L (1-2c) + (1-2c) \frac{\partial \chi(s)}{\partial s},
%\end{equation}
By substituting these $\rho(s)$ and $\chi(s)$ by~(\ref{Expectation_variance1}) and~(\ref{Expectation_variance2}), changing the variables to $x$, and using~(\ref{a*(s)})
and~(\ref{A_as_Function_Of_x}), we rewrite it as
\begin{eqnarray}
& - \frac{\partial \psi(s)}{\partial s} \big{|}_{s = s_c + \kappa^{-1} x} \nonumber \\
&  =  \frac{1}{2}\left [\left \langle \rho \right \rangle _i + \left \langle \rho \right \rangle_a 
\right ] \left (L c + c -1 \right ) +\frac{1-2c}{2}L \left [\left \langle \rho^2 \right \rangle _i + \left \langle \rho^2 \right \rangle_a 
\right ] \\
& + \frac{2x \left [\left \langle \rho_a \right \rangle -\left \langle \rho \right \rangle_i \right ]^{-1}}{\sqrt{1+4x^2 \left [\left \langle \rho_a \right \rangle -\left \langle \rho \right \rangle_i \right ]^{-2} }} \nonumber \\
&\times \left \{
\frac{1}{2}\left [\left \langle \rho \right \rangle _i - \left \langle \rho \right \rangle_a 
\right ] \left (L c + c -1 \right ) +\frac{1-2c}{2}L \left [\left \langle \rho^2 \right \rangle _i - \left \langle \rho^2 \right \rangle_a 
\right ].
\right \}
\end{eqnarray}
Finally, from this expression, we arrive at the asymptotic expression of $\partial \psi(s) / \partial s $ and $\partial^2 \psi(s) / \partial s^2$ as
\begin{equation}
-\lim_{L\rightarrow \infty} \frac{1}{L}\frac{\partial \psi(s)}{\partial s} \bigg{|}_{s = s_c + \kappa^{-1} x} = c^2 (1-c) \left [1 - \frac{2 x c^{-1}}{\sqrt{1+4 x^2 c^{-2}}} \right ],
\label{eq:scaling-psiprime}
\end{equation}
\begin{equation}
\lim_{L\rightarrow \infty} \frac{1}{L\kappa}\frac{\partial^2 \psi(s)}{\partial s^2} \bigg{|}_{s = s_c + \kappa^{-1} x} = 2c(1-c)\frac{1}{(1+4x^2c^{-2})^{3/2}}.
\label{eq:scaling-psiprimeprime}
\end{equation}
The results of this subsection are directly related to the ones~(\ref{analyscalingfunction1_infinite}) and~(\ref{analyscalingfunction2_infinite}) on the density and variance of occupied sites, and we believe that this connection is generic. The form~\eqref{eq:scaling-psiprime} and~\eqref{eq:scaling-psiprimeprime} provide us a complete description of the fluctuations of the dynamical activity around the transition point, where the fluctuations of the activity are at the origin of the transition itself.

%In the figure ..., we observe the convergence of exact results to these analytical expressions. 

%(Infinite L limit
%Infinite time limit
%Notion of s-state. Solving $\rho(s,t')$ for $\rho(0)=c$ and $0<t'<t$. 
%Evolution operator, spin representation, symmetrized version, related path integral.)

\section{Mean-field quantum ferromagnet and the scaling function}
\label{sec:MFqf}
In this section, we apply the method in the previous section to
a mean-field quantum ferromagnet.
We will show that this system has the same scaling functions as~(\ref{analyscalingfunction1}) and~(\ref{analyscalingfunction2}).
Furthermore, by applying our method to this model, we re-derive a formula that gives the exponent of an exponentially small gap at the quantum phase transition point~\cite{1742-5468-2012-06-P06007}.

\subsection{Set up}
\label{Sec/MFqf/Setup}
Let us consider~$L$ interacting 1/2 spins.
The Hilbert space is spanned by the space
$\{ |\vec{\sigma}\rangle \:|\: \vec{\sigma}=(\sigma_1,\cdots,\sigma_L)\in \{-1,+1\}^N\}$. We denote
the Pauli matrices acting on the $i$-th spin by
$\hat \sigma^x_i$, $\hat \sigma^y_i$, and $\hat \sigma^z_i$. These matrices satisfy
$\sigma_i^z |\vec{\sigma}\rangle =  \sigma_i |\vec{\sigma}\rangle$,
$\sigma_i^x |\vec{\sigma}\rangle =  |\vec{\sigma}^{(i)}\rangle$, where
$\vec{\sigma}^{(i)}$ is the configuration in which the $i$-th spin is flipped.
The transverse and the longitudinal magnetizations are defined as
\begin{equation}
\hat m^x=\frac{1}{L}\sum_{i=1}^L\hat \sigma_i^{x},
\end{equation}
\begin{equation}
\hat m^z=\frac{1}{L}\sum_{i=1}^L\hat \sigma_i^{z}.
\end{equation}
The Hamiltonian of the mean-field $p$-spin ferromagnet is defined as
\begin{equation}
\hat H =-L(\hat m^z)^p - \Gamma L \hat m^x
\label{Hamiltonian_Pspin_Ferromagnet}
\end{equation}
There is a phase transition in this model for a special value of $\Gamma$. 
It is known that for the $p=2$ (quantum Curie-Weiss model) the transition is second-order, whereas for the $p\geq 3$ the transition is first-order. See Ref.~\cite{1742-5468-2012-06-P06007} for the details of the thermodynamic properties of this model.

Here, we discuss the eigenvalues of the Hamiltonian $\hat H$. We denote by $| \Phi \rangle$
the eigenstate, and by $E$ the eigenvalue.
We especially focus on the eigenstates, where the interchanges of two spins are permitted. In other words,
we focus on the eigenstates that only depend on $m^z=(1/L)\sum_{i=1}^{L}\sigma_i^{z}$:
\begin{equation}
\langle \vec{\sigma}  |\Phi \rangle  = \Phi(m^z) .
\label{sigma_symmetric}
\end{equation}
We note that the ground state of
$\hat H$ lies in this symmetric subspace. 
See Ref.~\cite{1742-5468-2012-06-P06007} for the proof.
By multiplying the eigenvalue equation $\hat H| \Phi \rangle = E | \Phi \rangle $ by $\langle \vec{\sigma} |$ from the left, and using~(\ref{sigma_symmetric}), we obtain
%\begin{equation}
% -L( m^z)^p \langle \vec{\sigma}  |\Phi \rangle - \Gamma L  \frac{1}{L}\sum_{i=1}^{L}
% \langle \vec{\sigma}^{(i)}   | \Phi \rangle = E \langle \vec{\sigma}  |\Phi \rangle.
% \label{quantumferro_setup1}
%\end{equation}
%From (\ref{sigma_symmetric}) we rewrite 
%(\ref{quantumferro_setup1}) as
\begin{eqnarray}
& -L( m^z)^p \Phi(m^z) - \Gamma L  
 \frac{1}{L}\sum_{i=1}^{L}\left [ \frac{1+\sigma_i^{z}}{2}  \Phi(m^z-2/L)+ \frac{1 - \sigma_i^{z}}{2} \Phi(m^z+2/L) \right ] \nonumber \\
& = E \Phi(m^z),
\end{eqnarray}
which leads to an eigenvalue equation for the symmetric space,
\begin{equation}
\sum_{m^{\prime}\in \mathcal M}H_{m,m^{\prime}}\Phi(m^{\prime})
 = \frac{E}{L} \Phi(m),
 \label{Eigenvalue_Quantum}
\end{equation}
where $\mathcal M=\{-1,-1+2/L,\cdots,1-2/L,1 \}$ and
\begin{equation}
H_{m,m^{\prime}}
= -( m)^p \delta_{m,m^{\prime}} - \Gamma  
 \left [ \frac{1+m}{2} \delta_{m-2/L,m^{\prime}}+ \frac{1 - m}{2} \delta_{m+2/L,m^{\prime}} \right ].
 \label{Definition_Hmm}
\end{equation}
Thanks to the symmetry of the eigenstate, the dimension of the eigenvalue problem is reduced to $L+1$. 
We note that the matrix $H_{m,m^{\prime}}$ is not symmetric although the Hamiltonian 
$\hat H$ is Hermitian.
Here, we define the number of the state $p(m)$ by
\begin{equation}
p(m)\equiv \sum_{\vec{\sigma}}\delta_{m^z(\vec{\sigma}),m}=\frac{L!}{((1+m)L/2)!((1-m)L/2)!}\frac{1}{2^L}.
\label{numberofthestate}
\end{equation}
With this function, we may calculate the expectation value of a function $g(\hat m^{z})$
in these symmetric eigenstates from
\begin{equation}
\frac{\langle \Phi |  g(\hat m^z)  | \Phi \rangle}{\langle \Phi | \Phi \rangle} = \sum_{m} g(m) p^{\Gamma}(m),
\end{equation}
where the distribution function $p^{\Gamma}(m)$ is defined as
\begin{equation}
p^{\Gamma}(m)=\frac{p(m)|\Phi(m)|^2}{\sum_{m}p(m)|\Phi(m)|^2}.
\label{pGamma_def}
\end{equation}

Hereafter, we focus on the ground state of the system.
The expectation value and the variance of $\hat m^z$ for the ground state is also denoted by $m(\Gamma)$ and $\sigma(\Gamma)$,
which are calculated as
\begin{equation}
m(\Gamma)=\sum_{m} m p^{\Gamma}(m),
\label{m(Gamma)Definition}
\end{equation}
and 
\begin{equation}
\sigma(\Gamma)= L \sum_{m} (m - m(\Gamma))^2 p^{\Gamma}(m).
\label{sigma(Gamma)Definition}
\end{equation}
The ground state is determined by a variational principle.
That is, the ground state energy  
$E$ satisfies
\begin{equation}
E=\min_{|\Psi \rangle}\frac{\langle \Psi | \hat H  | \Psi \rangle}{\langle \Psi | \Psi \rangle}, 
\label{GroundState_VariationalPrinciple1}
\end{equation}
where the optimum is reached at the ground state $| \Phi \rangle$.
Because the ground state is in the symmetric space~\cite{1742-5468-2012-06-P06007}, 
we know that $|\Phi \rangle$ satisfies~(\ref{sigma_symmetric}). This allows us to rewrite 
(\ref{GroundState_VariationalPrinciple1}) as
\begin{equation}
\frac{E}{L}=\min_{\tilde \Phi>0} \sum_{m} \tilde p(m) \sum_{m^{\prime}} \tilde \Phi(m)^{-1}H_{m,m^{\prime}} \tilde \Phi (m^{\prime}), 
\label{GroundState_VariationalPrinciple2}
\end{equation}
where $\tilde p(m)$ is defined as
\begin{equation}
\tilde p(m) = \frac{p(m)\tilde \Phi(m)^2}{\sum_{m}p(m)\tilde \Phi(m)^2}.
\label{p_prime}
\end{equation}
The optimal $\tilde p$ is equal to the ground state distribution function $p^\Gamma$. The variational principle has the same structure as
(\ref{variationalprinciple2}). The modified distribution for the FA model~(\ref{Peq_prime}) (or~(\ref{distributionfunction_s}))
corresponds to the ground state of the quantum system~(\ref{p_prime}) (or (\ref{pGamma_def})).
This correspondence indicates that we might use the same technique as previously to approach the
finite-size properties for the quantum system --~which we indeed implement in the following subsections.

\subsection{Results in infinite-size limit}
\label{Sec/MFqf/ResultsInInfiniteSizeLimit}

By assuming the large deviation property for the ground state, we first
show a variational principle for the ground state energy in $L \rightarrow \infty$ and
determine the magnetization and the transverse field corresponding to the first order phase transition.

In the variational principle~(\ref{variationalprinciple2}), we assume a large deviation principle for
$\tilde p(m)$: $\tilde p(m)\sim \ee^{-L \tilde f(m)}$ with a large deviation function $\tilde f(m)$.
This indicates that $\tilde \Phi(m)$ also satisfies $\tilde \Phi \sim \ee^{-L\tilde \phi (m)/2}$ with a large deviation function $\tilde \phi(m)$.
From the large deviation property of $p(m)$, we have the relationship between $\tilde f(m)$
and $\tilde \phi(m)$ as
\begin{equation}
\tilde f(m)= \tilde \phi(m)+\frac{1+m}{2}\log(1+m)+\frac{1-m}{2}\log(1-m).
\end{equation}
The saddle point equation for $m$ is $\partial \tilde f(m)/\partial m=0$, which leads to
\begin{equation}
\frac{\partial \tilde \phi(m)}{\partial m}=\frac{1}{2}\log \frac{1+m}{1-m}.
\label{m*condition}
\end{equation}
By evaluating the variational principle~(\ref{variationalprinciple2}) with the saddle
point approximation, and using~(\ref{m*condition}) in it, we obtain
\begin{equation}
\frac{E}{L}=\min_{m} \left [-m^p - \Gamma \sqrt{1-m^2} \right ].
\label{Mean-Fieldvariational}
\end{equation} 
This variational formula is well-known. See Ref.~\cite{1742-5468-2012-06-P06007}, for example.
By solving this variational formula, we obtain an equation determining the expectation value 
of the magnetization $m^*$, which is
\begin{equation}
m^* \Gamma = p \left ( m^*\right )^{p-1} 
\sqrt{ 1-\left ( m^* \right )^2}.x
\label{MagnetInfinite}
\end{equation}
For $p\geq 3$, the system has the first order phase transition~\cite{1742-5468-2012-06-P06007} with a special value of $\Gamma$, which we denote $\Gamma_{\rm c}^{\infty}$. At the transition point, there are two solutions to the variational problem
(\ref{Mean-Fieldvariational}), corresponding to the paramagnetic solution $m^{\rm pa}_{\infty}=0$ and the ferromagnetic solution $m^{\rm fe}_{\infty}$. $\Gamma_{\rm c}^{\infty}$ and $m^{\rm fe}_{\infty}$ are determined from the conditions
\begin{equation}
\left [-m^p - \Gamma_{\rm c}^{\infty} \sqrt{1-m^2} \right ]\bigg |_{m^{\rm pa}_{\infty}=0} =
\left [-m^p - \Gamma_{\rm c}^{\infty} \sqrt{1-m^2} \right ]\bigg |_{m=m^{\rm fe}_{\infty}}
\label{Condition_Gammainf} 
\end{equation}
and~(\ref{MagnetInfinite}) with the replacement of $\Gamma$ and $m^*$ by $\Gamma_{\rm c}^{\infty}$ and $m^{\rm fe}_{\infty}$.

\begin{figure}[!h]
\centering
\includegraphics[width=9cm ]{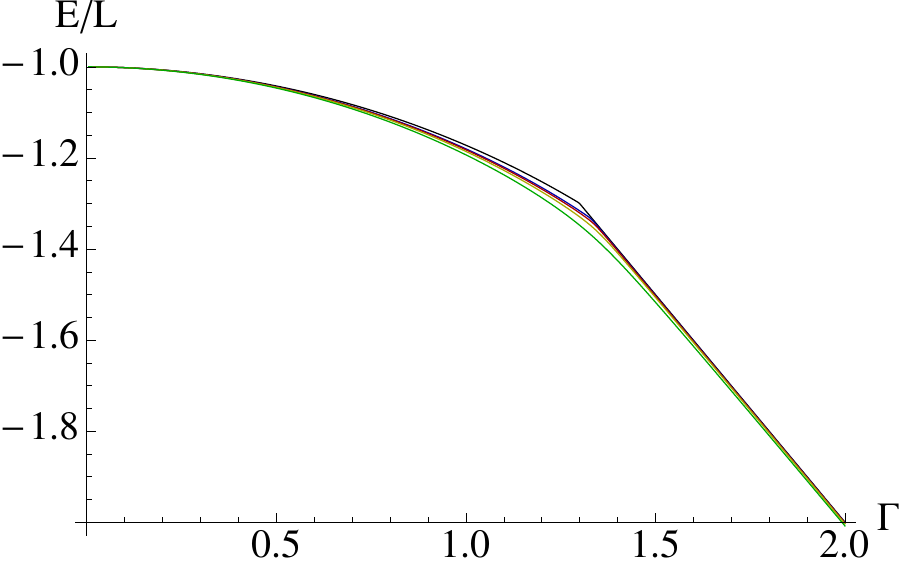}
\includegraphics[width=9cm ]{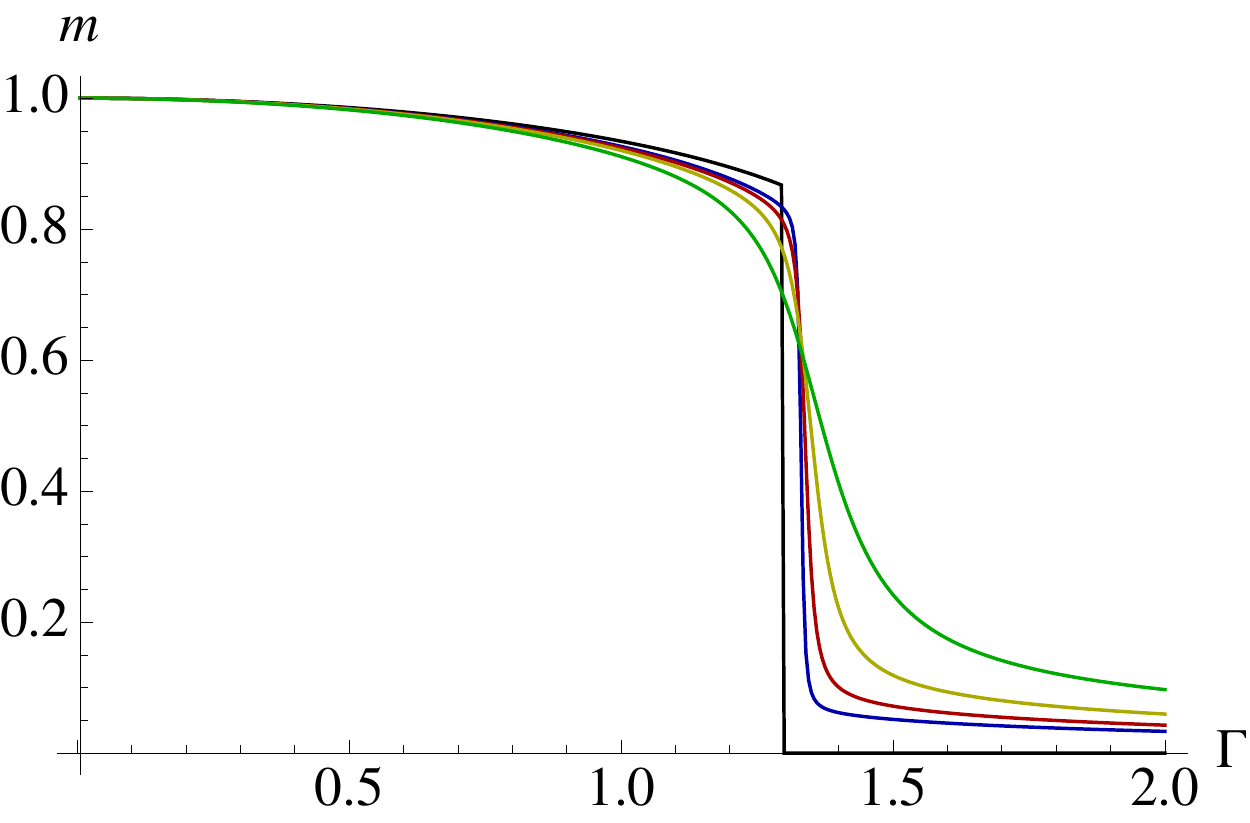}
\includegraphics[width=9cm ]{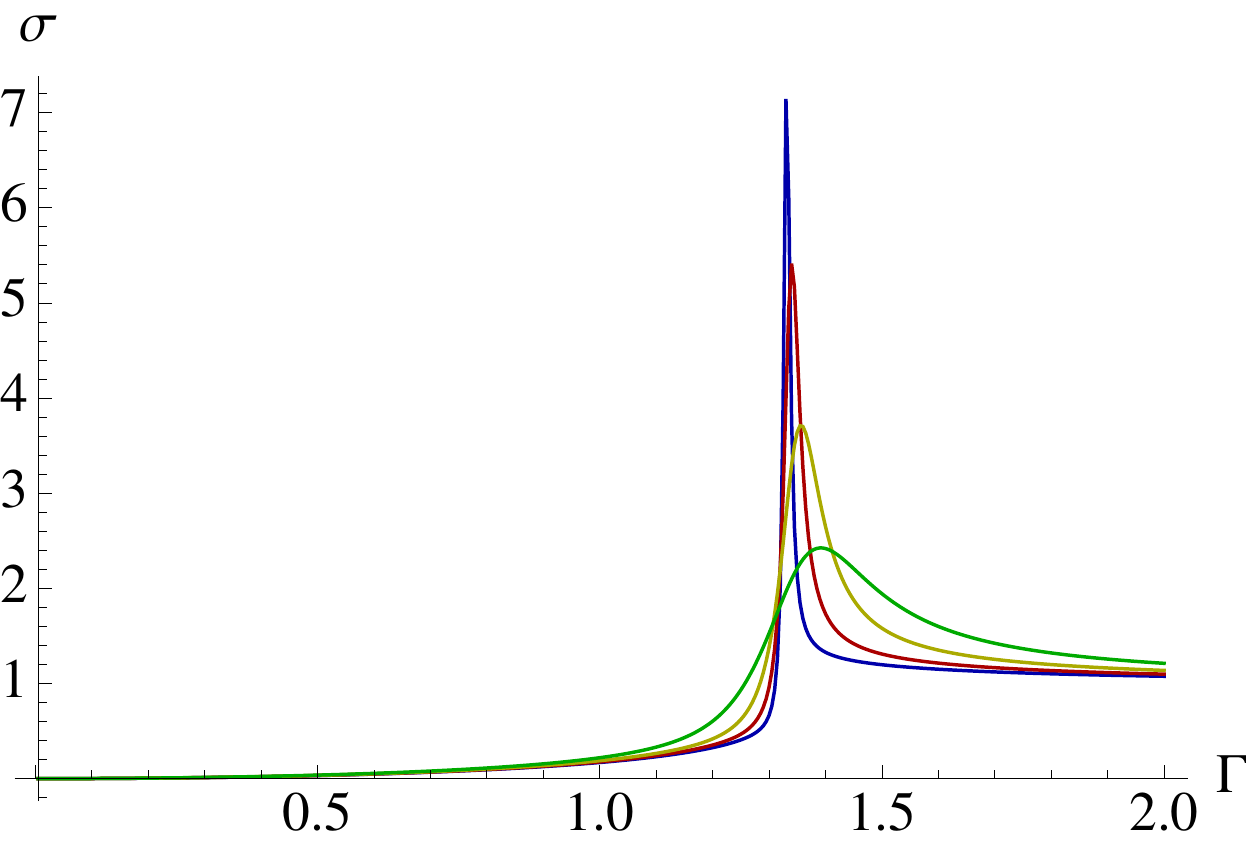}
\caption{\label{Figferro_groundenergy}The ground state energy $E/L$ (the black line on the upper figure) and the corresponding $m^*$ (the black line on the center figure) in the variational principle~(\ref{Mean-Fieldvariational}). At the same time, we also plot $E/L$ (upper), $m(\Gamma)$ (center) and $\sigma(\Gamma)$ (lower) obtained by the numerical diagonalization of the matrix~(\ref{Definition_Hmm}).
\textcolor{darkblue}{Blue}, \textcolor{darkred}{red}, \textcolor{darkyellow}{yellow} and \textcolor{darkgreen}{green} lines correspond to $L=50,40,30$, and $20$, respectively.
}
\end{figure}

In Fig.~\ref{Figferro_groundenergy},
we plot numerical examples of
$E/L$ and of the corresponding optimal~$m$ for $p=3$ obtained from the variational principle
(\ref{Mean-Fieldvariational}). At the same time, we also plot the numerical examples
of $E/L$, $m(\Gamma)$, and $\sigma(\Gamma)$ obtained from the direct diagonalization of 
the matrix~(\ref{Definition_Hmm}) for finite-size systems.
We can see the fist order phase transition around $\Gamma_{\rm c}^{\infty} \simeq 1.3$ in the figure. Also, we can see the finite-size correction of magnetization and susceptibility, which is the next target we consider.

\subsection{finite-size structure --~scaling functions}
\label{Sec/MFqf/ResultsFiniteSizeLimit}

\begin{figure}[h]
\centering
\includegraphics[width=9cm ]{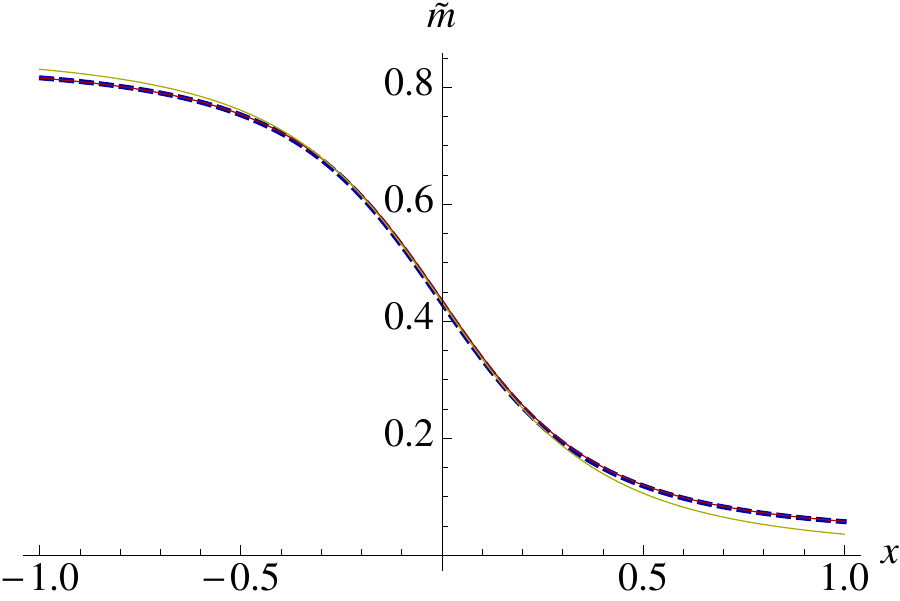}
\includegraphics[width=9cm ]{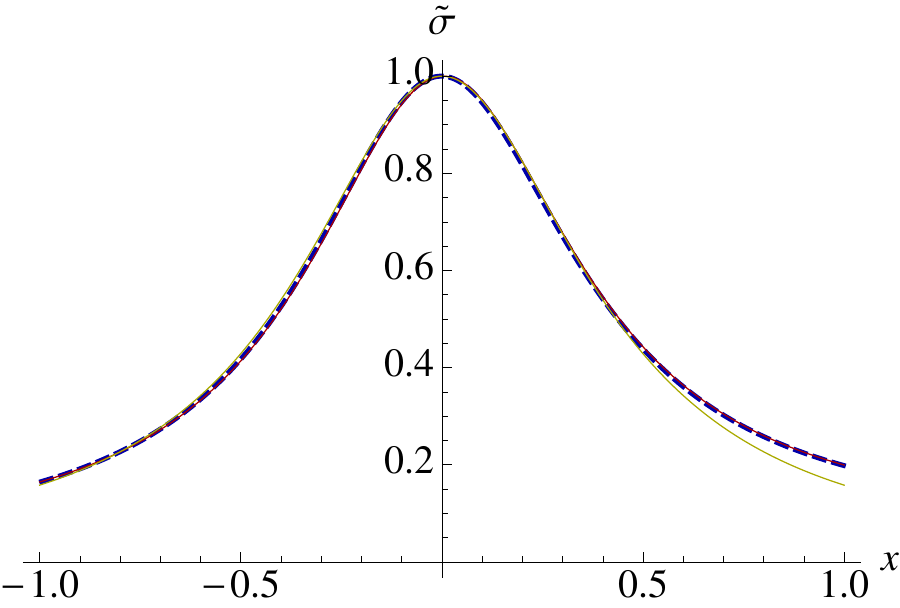}
\caption{\label{Figferro_scalingfunction}The scaling functions $\tilde m(x)$ (up)
and $\tilde \sigma(x)$ (down) for $p=3$ and $L=100$. The solid {\textcolor{darkred}{red lines}} are the analytical results~(\ref{analyscalingfunction_ferro_1}) and~(\ref{analyscalingfunction_ferro_2}). The dashed {\textcolor{darkblue}{blue lines}} are the numerical results
obtained from the direct diagonalization of~(\ref{Definition_Hmm}) for each $x$ (or $\Gamma)$.
For~(\ref{analyscalingfunction_ferro_1}) and~(\ref{analyscalingfunction_ferro_2}),
we need the distribution function $P_{\rm P}(m)$ and $P_{\rm f}(m)$ at $\Gamma=\Gamma_{\rm c}^{L}$. To obtain the distribution function, here, we evaluated
the corresponding eigenvector for $x=0$ (or $\Gamma=\Gamma_{\rm c}^{L}$), numerically. We also plot the infinite-size scaling functions~(\ref{analyscalingfunction_ferro_1_infinite}) and~(\ref{analyscalingfunction_ferro_2_infinite}) as solid {\textcolor{darkyellow}{yellow lines}}.}
\end{figure}

Now, we ask how to determine the finite-size structure shown in Fig.~\ref{Figferro_groundenergy}. For this purpose, we introduce the transition point $\Gamma_\rc^L$ for the finite-size system as
\begin{equation}
\Gamma_\rc^L={\rm Argmax}_{\Gamma}\sigma(\Gamma),
\label{Gamma_c^LDefinition}
\end{equation} 
which depends on~$L$. Then we define the
scaling ratio $\kappa$ by $-\partial m(\Gamma_\rc^L)/\partial \Gamma$. By using these quantities, we 
define
scaling functions
\begin{equation}
\tilde m(x) = m(\Gamma_\rc^L+x\kappa^{-1}),
\label{tildem(x)definition}
\end{equation}
and
\begin{equation}
\label{tildesigma(x)definition}
\tilde \sigma(x) = \frac{\sigma(\Gamma_\rc^L+x\kappa^{-1})}{\sigma(\Gamma_\rc^L)}.
\end{equation}
The question is how we determine the analytical expression of those scaling functions.
For this, we apply the same method as the previous section for the KCM.
First, we consider the distribution function $p^{\Gamma}(m)$ at the transition point.
We assume that the distribution function is divided into two regions, the paramagnetic
region $P_{\rm p}(m)$ and the ferromagnetic region $P_{\rm f}(m)$:
\begin{equation}
p^{\Gamma_\rc^L}(m)=\delta_{m\leq m_\rc}P_{\rm p}(m)
+\delta_{m>m_\rc}P_{\rm f}(m),
\label{pGamma_ConnectingAnsatz}
\end{equation}
where $m_\rc$ is the boundary of these two regions, which may be defined as
the valley between two peaks for $\log P_{\rm p}(m)$.
We note that $P_{\rm p}(m)$ and $P_{\rm f}(m)$ satisfy the condition of first-order phase transitions
\begin{equation}
\sum _{m\leq m_\rc}P_{\rm p}(m) = \sum _{m > m_\rc} P_{\rm f}(m) = 1/2.
\label{1stTransition_Ferro}
\end{equation}
Next, for the region around $\Gamma_\rc^L$, we assume that $p^{\Gamma}$ is written as
\begin{equation}
p^{\Gamma}(m)=(1+ a^*(\Gamma))\delta_{m\le m_\rc}P_{\rm p}(m)
+(1-a^*(\Gamma))\delta_{m>m_\rc}P_{\rm f}(m),
\label{AssumptionpGamma_aroundGammac}
\end{equation}
where $a^*(\Gamma)$ is a mixing function of $\Gamma$. We note that the normalization condition is satisfied due 
to 
(\ref{1stTransition_Ferro}).
From the distribution function, we can calculate $m(\Gamma)$, $\sigma(\Gamma)$ as
\begin{equation}
m(\Gamma) = \frac{\left \langle m \right \rangle_{\rm p}}{2}\left [1+a^*(\Gamma) \right ]
+ \frac{\left \langle m \right \rangle_{\rm f}}{2}\left [1-a^*(\Gamma) \right ],
\label{m_around_transitionpoint}
\end{equation}
\begin{equation}
\sigma(\Gamma) = L\left \{
\frac{\left \langle m^2 \right \rangle_{\rm p}}{2}\left [1+a^*(\Gamma) \right ]
+ \frac{\left \langle m^2 \right \rangle_{\rm f}}{2}\left [1-a^*(\Gamma) \right ] - m (\Gamma)^2 
\right \},
\label{sigma_around_transitionpoint}
\end{equation}
where $\left \langle \ \right \rangle_{\rm p}$ and $\left \langle \ \right \rangle_{\rm f}$
are the expectation values in the paramagnetic phase and the ferromagnetic phase, respectively, which are
 defined as
$\left \langle g \right \rangle_{\rm p}=2 \sum_{m\leq m_\rc}P_{\rm p}(m) g(m)$ and $\left \langle g \right \rangle_{\rm f}=2 \sum_{m > m_\rc}P_{\rm f}(m) g(m)$.
For the determination of $a^*(\Gamma)$, we use the variational principle~(\ref{GroundState_VariationalPrinciple2}). Indeed, from~(\ref{AssumptionpGamma_aroundGammac}) with
(\ref{pGamma_def}), we have $\Phi(m)$. By substituting the obtained $\Phi(m)|_{a^*(\Gamma)=a}$ and $p^{\Gamma}(m)|_{a^*(\Gamma)=a}$
into the variational functional of~(\ref{GroundState_VariationalPrinciple2}) and maximizing
it with respect to $a$, we obtain the optimal $a^*$, which corresponds to $a^*(\Gamma)$.
The result is
\begin{equation}
a^*(\Gamma) = \frac{x \left [ \left \langle m \right \rangle_{\rm p}
- \left \langle m \right \rangle_{\rm f}\right ]^{-1} }{\sqrt{1+4 x^2\left [ \left \langle m \right \rangle_{\rm p}
- \left \langle m \right \rangle_{\rm f}\right ]^{-2}}} + O(\kappa^{-1}).
\end{equation}
with $x=\kappa (\Gamma - \Gamma_\rc)$.
Combining it with~(\ref{m_around_transitionpoint}) and~(\ref{sigma_around_transitionpoint}), 
we obtain
\begin{equation}
\tilde m(x)=\frac{1}{2} \left [
\left \langle m \right \rangle_{\rm p}
+\left \langle m \right \rangle_{\rm f}
- \frac{2x}{\sqrt{1+4x^2\left [\left \langle m \right \rangle_{\rm p}
-\left \langle m \right \rangle_{\rm f} \right ]^{-2} }}
\right ],
\label{analyscalingfunction_ferro_1}
\end{equation}
\begin{eqnarray}
&\tilde \sigma(x) =\nonumber \\
&\;\frac{1}{C}
\left [
\left \langle m ^2\right \rangle_{\rm p}
+\left \langle m^2\right \rangle_{\rm f}
-   \frac{2x \left [\left \langle m ^2\right \rangle_{\rm p}
-\left \langle m^2\right \rangle_{\rm f} \right ]
 \left [\left \langle  m \right \rangle_{\rm p}
-\left \langle  m \right \rangle_{\rm f} \right ]^{-1}}
{\sqrt{ 1+4x^2\left [\left \langle  m \right \rangle_{\rm p}
-\left \langle m \right \rangle_{\rm f} \right ]^{-2} }} - 2 \tilde m(x)^2 \right ],
\label{analyscalingfunction_ferro_2}
\end{eqnarray}
where $C$ is
\begin{equation}
C=\left \langle m ^2\right \rangle_{\rm p}
+\left \langle m^2\right \rangle_{\rm f}
 - 2 \tilde m(0)^2.
\end{equation}
These expressions are equivalent to the relations
(\ref{analyscalingfunction1})
and~(\ref{analyscalingfunction2}) that we obtained in our study of the KCM. 
Noticing that $\lim_{L \rightarrow \infty} \left \langle m \right \rangle_{\rm p}=\lim_{L \rightarrow \infty}  \left \langle m^2 \right \rangle_{\rm p}=0$, $\lim_{L \rightarrow \infty} \left \langle m \right \rangle_{\rm f}=m_{\infty}^{\rm fe}$, and $\lim_{L \rightarrow \infty}  \left \langle m^2 \right \rangle_{\rm f}=(m_{\infty}^{\rm fe})^2$, we also obtain the infinite-size scaling functions:
\begin{equation}
\tilde m_{\infty}(x)=  \lim_{L\rightarrow \infty} \tilde m(x) =\frac{1}{2}  \left [ m_{\infty}^{\rm fe} - \frac{2x}{\sqrt{1 + 4 x^2 (m_{\infty}^{\rm fe})^{-2} }} \right ],
\label{analyscalingfunction_ferro_1_infinite}
\end{equation}
\begin{equation}
\tilde \sigma_{\infty} (x)= \lim_{L\rightarrow \infty} \tilde \sigma(x) =\frac{(m_{\infty}^{\rm fe})^2}{(m_{\infty}^{\rm fe})^2+4 x^2},
\label{analyscalingfunction_ferro_2_infinite}
\end{equation}
which correspond to~(\ref{analyscalingfunction1_infinite})
and~(\ref{analyscalingfunction2_infinite}).
We check the obtained results in Fig.~\ref{Figferro_scalingfunction}, from which
one can see that~(\ref{analyscalingfunction_ferro_1}) and~(\ref{analyscalingfunction_ferro_2}) (solid red lines) show good agreement with the  numerical results from direct diagonalization (blue dotted lines). As the same as the previous section, we also note that large system sizes are required to observe the convergence to the infinite-size scaling functions~(\ref{analyscalingfunction_ferro_1_infinite}) and~(\ref{analyscalingfunction_ferro_2_infinite}).

\subsection{finite-size structure -- scaling factor and exponentially small gap}
\label{Sec/MFqf/scaling factor}

Finally, we derive the exponent of the scaling factor $\kappa$, which is equivalent to the exponentially small gap derived in~\cite{1742-5468-2012-06-P06007}.

We first define the free energy for the ground state at the transition point $\Gamma_{\rm c}^{\infty}$ by
\begin{equation}
f_{\Gamma_{\rm c}^{\infty}}(m)=-\lim_{L\rightarrow \infty}\frac{1}{L} \log p^{\Gamma_{\rm c}^{\infty}}(m).
\label{f_Gamma(m)Ferromagnet_Defition}
\end{equation}
Then, using  the same argument as for~(\ref{kapparesult}), we obtain the exponent $\kappa$
\begin{equation}
\lim_{L\rightarrow \infty} \frac{1}{L}\log \kappa = f_{\Gamma_{\rm c}^{\infty}}(m_{\rm c}^{\infty}),
\label{GapFormula}
\end{equation}
where $m_{\rm c}^{\infty} = \lim_{L\rightarrow \infty}m_{\rm c}$ represents the connecting point between the paramagnetic and the ferromagnetic regions in the infinite system size limit.

Next, we determine the free energy. We start from the ground state of the eigenvalue equation for $\Gamma=\Gamma_{\rm c}^{\infty}$, 
\begin{equation}
-m^{p} - \Gamma_{\rm c}^{\infty} \left [ \frac{1+m}{2}\frac{\Phi(m-2/L)}{\Phi(m)} + \frac{1-m}{2} \frac{\Phi(m+2/L)}{\Phi(m)} \right ] = \frac{E}{L},
\label{EigenQuantum_2}
\end{equation}
which is obtained from~(\ref{Eigenvalue_Quantum}). 
Now, we assume a large deviation principle. That is,
we set $\Phi(m)=e^{-(L/2)\phi(m)}$ in~(\ref{EigenQuantum_2}). The leading term of~(\ref{EigenQuantum_2}) is
\begin{equation}
-m^{p}-\Gamma_{\rm c}^{\infty} \left [ \frac{1+m}{2}e^{\partial \phi /\partial m} + \frac{1-m}{2} e^{- \partial \phi /\partial m} \right ] = e^{\infty}_{\rm c},
\label{e_cfirstapperared}
\end{equation}
where we define $\lim_{L\rightarrow \infty}E/L|_{\Gamma=\Gamma_{\rm c}^{\infty}} \equiv e^{\infty}_{\rm c}$.
By solving this equation, we obtain two solutions for the expression $\partial \phi(m)/\partial m$ as
\begin{eqnarray}
& \phi(m)  + {\rm const.} \nonumber \\
&= \phi_{\pm}(m) \nonumber \\
&\equiv \int_{0}^{m} d\tilde m
\log \left [ - \frac{\tilde m^{p}+e^{\infty}_{\rm c}}{(1+\tilde m)\Gamma_{\rm c}^{\infty}} \pm \sqrt{\left (\frac{\tilde m^{p}+e^{\infty}_{\rm c}}{(1+\tilde m)\Gamma_{\rm c}^{\infty}}\right )^2 -\frac{1-\tilde m}{1+\tilde m} } \right ]
\end{eqnarray}
By using these two functions, we also define free energies as
\begin{eqnarray}
f_{\pm}(m) & \equiv  \phi_{\pm}(m) + \frac{1+m}{2}\log (1+m) + \frac{1-m}{2}\log (1-m) \nonumber \\
&=\int_{0}^{m} d\tilde m
\log \left [ - \frac{\tilde m^{p}+e^{\infty}_{\rm c}}{\sqrt{1-\tilde m^2}\Gamma_{\rm c}^{\infty}} \pm \sqrt{\left (\frac{\tilde m^{p}+e^{\infty}_{\rm c}}{\sqrt{1-\tilde m^2}\Gamma_{\rm c}^{\infty}}\right )^2 - 1 } \right ]
\end{eqnarray}
The free energy $f_{\Gamma_{\rm c}^{\infty}}(m)$ is given as the combination of $f_{+}(m)$ and $f_{-}(m)$.
Here, we notice
\begin{equation}
f_{\Gamma_{\rm c}^{\infty}}(0) = f_{\Gamma_{\rm c}^{\infty}}(m_{\rm fe}^{\infty}) = 0.
\label{FirstOrderCondition}
\end{equation}
Also, we can easily check
\begin{equation}
\frac{\partial f_{+}(m)}{\partial m}\bigg | _{m=0} > 0
\label{f+deri0}
\end{equation}
\begin{equation}
 \frac{\partial f_{-}(m)}{\partial m}\bigg | _{m=0} < 0.
\label{f-deri0}
\end{equation}
From~(\ref{FirstOrderCondition}), (\ref{f+deri0}) and~(\ref{f-deri0}), we can construct the free energy $f_{\Gamma_{\rm c}^{\infty}}(m)$ as
\begin{equation}
f_{\Gamma_{\rm c}^{\infty}}(m) = f_{+}(m)
\end{equation}
for $m \leq m_{\rm c}^{\infty}$
\begin{equation}
f_{\Gamma_{\rm c}^{\infty}}(m) = f_{-}(m) + \rm const.
\end{equation}
for $m > m_{\rm c}^{\infty}$. The constant and $m_{\rm c}^{\infty}$ is determined from~(\ref{FirstOrderCondition}) and the continuity condition 
\begin{equation}
\lim_{m\rightarrow m_{\rm c} +0}f_{\Gamma_{\rm c}^{\infty}}(m)=\lim_{m\rightarrow m_{\rm c} -0}f_{\Gamma_{\rm c}^{\infty}}(m).
\end{equation}

By using the parameters $\Gamma_{\rm c}^{\infty}$, $e^{\infty}_{\rm c}$, and $m_{\rm fe}^{\infty}$ obtained from~(\ref{Mean-Fieldvariational}), (\ref{MagnetInfinite}) and~(\ref{Condition_Gammainf}), we can calculate the gap given by~(\ref{GapFormula}) in principle. Here, however, by using a relation
\begin{equation}
f_{+}(m) = - f_{-}(m) + \rm const.,
\end{equation}
which can be derived from the direct substitution, we will show that the gap can be expressed as a simple formula. Indeed, from this relation, we can derive
\begin{equation}
f_{+}(m_{\rm c}^{\infty}) = \frac{1}{2}f_{+}(m_{\rm fe}^{\infty}).
\end{equation}
Thus, we arrive at
\begin{eqnarray}
&\lim_{L\rightarrow \infty} \frac{1}{L}\log \kappa \nonumber \\ 
&= \frac{1}{2}f_{+}(m_{\rm fe}^{\infty}) \nonumber \\
&=\int_{0}^{m_{\rm fe}^{\infty}} d m
\log \left [ - \frac{ m^{p}+e^{\infty}_{\rm c}}{\sqrt{1-m^2}\Gamma_{\rm c}^{\infty}} + \sqrt{\left (\frac{ m^{p}+e^{\infty}_{\rm c}}{\sqrt{1-m^2}\Gamma_{\rm c}^{\infty}}\right )^2 -1 } \right ]
\end{eqnarray}
This formula is equivalent to the formula obtained by Bapst and Semerjian, Eq.~(62) in 
Ref.~\cite{1742-5468-2012-06-P06007}. In order to see the equivalence, we just use a basic mathematical fact that the following equations $\cosh x = A$ and $e^x=A\pm \sqrt{A^2-1}$ are equivalent.

\section{Conclusions}
Our goal in this work was to identify the characteristic features of finite-size scaling at a first-order dynamical transition, as can be found generically in KCMs. As we have shown, these are akin to characteristic features of a first-order \textit{quantum} transition. We have been able, in particular, to determine in an explicit fashion the scaling variables and the scaling functions governing the variation of the order parameter across the transition. For the particular mean-field KCM that we have been considering here, a precise characterization of a size-dependent critical point was provided and the finite-size rounding off of the transition was fully captured by our study. Interestingly, the phenomenology of our dynamical transition --~which is identical to that of quantum transitions, but now capturing finite-size scaling~-- also agrees with that of classical first order transitions, as our variational formulation in terms of the mixing function $a^*(s)$ confirms. 

We also note that after the submission of our work, Campostrini and collaborators presented results in a recent preprint~\cite{campostrini_finite-size_2014} on the finite-size scaling of first-order quantum phase transition. Close to the critical point, scaling functions for their order parameter take the same form as presented in our approach, with a derivation based on a two-level effective model. Further connections between our work and this approach (and also with the two-level effective model of~\cite{0295-5075-89-4-40004} used for an instantonic computation of the gap) are worth studying.

The mean-field version being now fully understood, 
a more challenging program awaits ahead of us. Finite-dimensional 
systems of course display a richer phenomenology~\cite{bodineautoninelli} with, for instance, the existence of surface tension and nucleating droplets. It is in principle possible to extend our analysis for the density large deviation function to a space varying field, by means of field theoretic methods {\it \`a la} Doi-Peliti~\cite{0305-4470-9-9-008,refId0}, for example. When studying dynamical transitions, some authors~\cite{hedges_dynamic_2009} work at fixed system size, but they perform a finite time analysis (which would be mimicked by a finite temperature analysis in quantum phase transition). It would certainly be of interest to quantify in a similar way finite-time corrections, though the phase transition itself is indeed a collective effect captured only at large~$N$. These issues are currently under investigation.

\medskip\noindent
\textbf{Acknowledgments} -- we would like to thank an anonymous referee for pointing out an error 
in a previous version of subsection~\ref{roughs_c}.
This work was supported by JSPS Core-to-Core program
``Non-equilibrium dynamics of soft matter and information'' and by the ``LaBS'' PEPS CNRS project.

\addcontentsline{toc}{section}{Appendices}
\begin{appendices}
\section{Determination of $s_c$}
\label{app:sc_expression_derivation}

%We evaluate (\ref{Conditionfors_c}) by using explicit expression of $r$, $\tilde r_{\rm i}$, and $\tilde r_{\rm a}$. From the large deviation property, we evaluate $\cdots$

In this appendix, we derive $s_\rc=1/(2Lc(1-c))+O(1/L^2)$.
We evaluate~(\ref{Conditionfors_c}) up to $O(1/L)$ by using the explicit expressions of $r$, $\tilde r_{\rm i}$, and $\tilde r_{\rm a}$. 
From a saddle point approximation, we rewrite the left-hand side of~(\ref{Conditionfors_c}) as
\begin{equation}
\frac{1}{L} \left ( \tilde r_{\rm i}\ee^{-s_\rc} - r  \right ) \Big | _{n=1} +O(1/L^2) = \frac{c}{L} \ee^{-s_\rc}\frac{\ee^{-L f_{\rm i}(2/L)/2} }{\ee^{-L f_{\rm i}(1/L)/2} } - \frac{c}{L}  + O(1/L^2).
\end{equation}
Here, the first term is $O(1/L^2)$ since 
\begin{equation}
\frac{\ee^{-L f_{\rm i}(2/L)/2} }{\ee^{-L f_{\rm i}(1/L)/2} } = O(1/L).
\end{equation}
Thus, the left-hand side of~(\ref{Conditionfors_c}) is $-c/L + O(1/L^2)$. On the other hand, the right-hand side of~(\ref{Conditionfors_c}) is evaluated by saddle point approximation as
\begin{eqnarray}
&\frac{1}{L}\left ( \tilde r_{\rm a}\ee^{-s_\rc} - r  \right ) \Big |_{n=Lc} + O(1/L^2) \nonumber \\
&= c^2(1-c)\left [\ee^{-\frac{1}{2} \frac{\partial f_a(\rho)}{\partial \rho}|_{\rho = c} -s_\rc } + \ee^{\frac{1}{2} \frac{\partial f_a(\rho)}{\partial \rho}|_{\rho = c} -s_\rc }  -2 \right ] +O(1/L^2) .
\label{Derivationsc1}
\end{eqnarray}
By noticing that $\partial f_a(\rho) / \partial \rho|_{\rho = c}$ and $s_\rc$ are $O(1/L)$, we rewrite~(\ref{Derivationsc1}) as
\begin{equation}
-2s_\rc\, c^2(1-c) +O(1/L^2).
\end{equation}
Therefore, by equating the left-hand side of~(\ref{Conditionfors_c}) to the right-hand side of~(\ref{Conditionfors_c}), we arrive at
\begin{equation}
s_\rc = \frac{1}{2Lc(1-c)} + O(1/L^2).
\end{equation}

\section{Finite-size corrections to the free energy difference}
\label{app:Derivation_approximation}

In this appendix, we derive the finite-size correction $\Delta f_{s_{\rm c}}^{(1)}(\rho)$ given in~(\ref{deltafsmall}) and~(\ref{deltaflarge}).
We first focus on the region $\rho > \rho_\rc^{L}$. From~(\ref{EigenEquation_Derivation}) with~(\ref{Def_Deltaf1}), we find that
$\Phi_{\rm L}(n)$ doesn't satisfy the large deviation principle.
Then, we define $\tilde \Phi_{\rm L}(\rho) = \Phi_{\rm L}(\rho L)$. From the fact $\Phi_{\rm L}(n)$ doesn't satisfy the large deviation scaling, we can assume $\tilde \Phi_{\rm L}(\rho)$ is differentiable:
\begin{equation}
\tilde \Phi_{\rm L}(\rho \pm 1/L) = \tilde \Phi_{\rm L}(\rho) \pm \frac{\partial \tilde \Phi_{\rm L} }{\partial \rho} \frac{1}{L}+O(1/L^2).
\label{differentiability_phi}
\end{equation}
By rewriting the left-hand side of~(\ref{EigenEquation_Derivation}) by using this scaling, we obtain a differential equation for determining $\tilde \Phi_{\rm L}(\rho)$.
\begin{equation}
\tilde \Phi_{\rm L}(\rho)\left \{ -\tilde s_\rc \rho \left [c+(1-2c)\rho \right ]+c\right \}+\frac{\partial \tilde \Phi_{\rm L}(\rho)}{\partial \rho}\rho(c-\rho)+O(1/L^2)=0,
\end{equation}
where we defined $\tilde s_c\equiv  sL$. By solving this differential equation,
we obtain
\begin{eqnarray}
&-2\log \tilde  \Phi_{\rm L}(\rho) \nonumber \\
&=-2\left [ \tilde s_\rc\rho (2c-1)-\log \rho+ (-\tilde s_\rc2c(1-c)+1)\log|c-\rho|\right ]+\rm const.
\end{eqnarray}
Finally with a relation $\tilde s_\rc = 1/(2c(1-c)) +O(1/L)$, we can omit the last term. 
The result leads to
\begin{eqnarray}
-2\log \tilde \Phi_{\rm L}(\rho) =\Delta f_{s_{\rm c}}^{(1)}(\rho)=-2\left [\frac{\rho(2c - 1)}{2c(1-c)} -  \log\rho\right ] + \rm const.
\label{regime2}
\end{eqnarray}

Next, we focus on the region $\rho \leq \rho_\rc^{L}$. By using the explicit expression of $\Delta f_{s_\rc}$ in this region, we evaluate  
the equation~(\ref{EigenEquation_Derivation}) up to $O(1/L)$.
First, $\Phi_{\rm L}(n+1)/\Phi_{\rm L}(n)$ is evaluated as
\begin{eqnarray}
\frac{\Phi_{\rm L}(n+1)}{\Phi_{\rm L}(n)} &= e^{\partial f_{\rm e}/\partial \rho +1/(2L)\partial ^2 f_{\rm e}/\partial \rho^2  -1/(2L) \partial \Delta f_{s_{\rc}}^{(1)}/\partial \rho} \nonumber \\ 
& = \frac{(1-c)\rho}{c(1-\rho)}e^{ (1/(2L\rho(1-\rho)))} e^{ -1/(2L) \partial \Delta f_{s_{\rc}}^{(1)}/\partial \rho}.
\end{eqnarray}
Then, we obtain an equation for determining $\Delta f_{s_{\rc}}^{(1)}$ as
\begin{eqnarray}
\frac{\partial \Delta f_{s_{\rc}}^{(1)} (\rho)}{\partial \rho} &= \frac{2}{\rho} - 2\frac{\frac{c}{\rho}+\left [ (1-c)\rho + c(1-\rho)\right ]
\left [\frac{1}{2\rho(1-\rho)} -\frac{1}{2c(1-c)}\right ]
}{-(1-c)\rho + c(1-\rho)}  \nonumber \\
&=-\frac{1}{\rho} +\frac{1}{1-\rho} - \frac{2}{c-\rho}  - \frac{1-2c}{c(1-c)},
\end{eqnarray}
which leads to
\begin{equation}
 \Delta f_{s_{\rc}}^{(1)} (\rho)
=-\log\frac{\rho (1-\rho)}{(c - \rho)^2} - \frac{\rho(1-2c)}{c(1-c)} + \rm const.
\end{equation}

\section{Table of notations}
\label{app:tableofnotations}
\begin{center}
  \begin{longtable}{rll}
%   \caption[Table of notations]{Table of notations}\\
   \textit{Quantity} & \textit{Notation} & \textit{Defining equation} 
   \\
   transition rates  & $w(n\rightarrow n')$ &
    \eqref{TransitionRateDefinition}
    \\
    equilibrium distribution  & $P_{\rm eq}(n)$ & 
    \eqref{eq:defPeq}
    \\
    equilibrium free energy   & $f_\eee(\rho)$ & 
    \eqref{eq:refIequilib}
    \\
    dynamical free energy or cumulant generating function & $\psi(s)$ &
    \eqref{eq:defpsis}
    \\
    matrix of evolution       & $W$ &
    \eqref{defmatrixL}
    \\
    left eigenvector of $W$ for the eigenvalue $L \psi(s)$ & $\Phi_{\rm L}$ &
    \eqref{largesteigenvalue}
    \\
    transition rates of the modified dynamics & $w_s(n\to n')$ &
    \eqref{eq:defws}
    \\
     equilibrium distribution for the modified dynamics &  $P^{s}(n)$ &
    \eqref{distributionfunction_s}
    \\
    free energy difference & $\Delta F_s(n)$ &
    \eqref{HamiDiff_def}
    \\
    finite-size critical point & $s_{\rm c}$ &
    \eqref{s_cDefinition}
    \\
    scaled critical point & $\lambda_{\rm c}$ &
    \eqref{lambda_cDefinition}
    \\
    rescaled dynamical free energy  & $\varphi _{L}(\lambda)$ &
    \eqref{rescaledFreeEnergy}
    \\
    density of free energy difference & $\Delta f_{\rm s}(\rho)$ &
    \eqref{DensityOfFreeEnergyDifference}
    \\
    infinite-size connecting point &$\rho_{\rm c}^{\infty}$&
    \eqref{equationForrhoc}      
    \\
    finite-size connecting point &$\rho_{\rm c}^{L}$&
    \eqref{condition_for_rhoc}      
    \\
    first order correction (in powers of $L$) to $\Delta f_{\rm s}(\rho)$ & $\Delta f^{(1)}_{\rm s}(\rho)$&
    \eqref{Def_Deltaf1}
    \\
    mean occupation number  & $\rho(s)$ &
    \eqref{eq:defrhos}
    \\
    variance of the occupation number  & $\chi(s)$ &
    \eqref{eq:defchis}
    \\
    scaling ratio & $\kappa$&
    \eqref{ScalingRationDef}
    \\
    scaling variable & $x$  &
    \eqref{definitionx}
    \\
    rescaled $\rho(s)$ & $\tilde \rho(x)$ &
    \eqref{RescaledMeanOccupation}
    \\
    rescaled $\chi(s)$ & $\tilde \chi (x)$  &
    \eqref{RescaleOccupationVariance}
    \\
    variational function for determining $\Phi_L$ & $\tilde \Phi_{\rm L}(n)$ &
    \eqref{variationalfunction_PhiL(n)}
    \\
   variational distribution & $\widetilde P(n)$ &
    \eqref{Peq_prime}
    \\
    variational function for the free energy difference & $\Delta \widetilde F (n)$ &
    \eqref{VariationalFreeEnergyDifference}
    \\
   variational function for the density of free energy difference & $\tilde f(\rho)$&
   \eqref{VariationalFunctionForDensityFreeEnergyDifference}
   \\
   finite-size free energy difference in inactive region & $f_{\rm i}(\rho)$&
   \eqref{f_iDefinition}   
   \\
   finite-size free energy difference in active region & $f_{\rm a}(\rho)$&
   \eqref{f_aDefinition}   
   \\
   distribution in inactive region at the transition point & $P_{\rm i}(n)$&
   \eqref{P_iDefinition}   
   \\
   distribution in active region at the transition point & $P_{\rm a}(n)$&
   \eqref{P_aDefinition}   
   \\
   mixing function & $a^{*}(s)$&
   \eqref{Ansatz1}   
   \\
   variational function with mixing parameter $a$ & $\Psi(a)$&
   \eqref{VariationalfunctionPsi}
   \\
   infinite-size scaling function of $\tilde \rho(x)$ &$\tilde \rho_{\infty}(x)$&
   \eqref{analyscalingfunction1_infinite}   
   \\
   infinite-size scaling function of $\tilde \chi(x)$ & $\tilde \chi_{\infty}(x)$&
   \eqref{analyscalingfunction2_infinite}   
   \\
   Hamiltonian of the mean-field $p$-spin ferromagnet & $\hat H$&
   \eqref{Hamiltonian_Pspin_Ferromagnet}      
    \\
    eigenvector of the Hamiltonian with interchange symmetry & $\Phi(m^z)$&
   \eqref{sigma_symmetric}      
    \\
    number of the state & $p(m)$&
   \eqref{numberofthestate}
    \\
    distribution of magnetization with transverse field $\Gamma$ & $p^{\Gamma}(m)$&
   \eqref{pGamma_def}
    \\
    ground state mean magnetization with transverse field $\Gamma$ & $m(\Gamma)$&
   \eqref{m(Gamma)Definition}      
    \\
    ground state susceptibility with transverse field $\Gamma$ & $\sigma(\Gamma)$&
   \eqref{sigma(Gamma)Definition}      
    \\
    transverse field at the transition point in infinite-size limit & $\Gamma_{\rm c}^{\infty}$&
    \eqref{Condition_Gammainf}      
    \\
    ground state mean magnetization at $\Gamma=\Gamma_{\rm c}^{\infty}$ in infinite-size limit &$m_{\infty}^{\rm fe}$&
    \eqref{Condition_Gammainf}      
    \\
    transverse field at the transition point for a finite-size system & $\Gamma_{\rm c}^{L}$&
    \eqref{Gamma_c^LDefinition}      
    \\
    rescaled $m(\Gamma)$ & $\tilde m(x)$&
   \eqref{tildem(x)definition}      
    \\
   rescaled $\sigma (\Gamma)$ & $\tilde \sigma(x)$&
   \eqref{tildesigma(x)definition}      
    \\
   finite-size connecting point & $m_{\rm c}$&
   \eqref{pGamma_ConnectingAnsatz}
    \\
   mixing function for the case of quantum ferromagnet & $a^*(\Gamma)$&
   \eqref{AssumptionpGamma_aroundGammac}      
    \\
   infinite-size scaling function for $\tilde m(x)$ & $\tilde m_{\infty}(x)$&
   \eqref{analyscalingfunction_ferro_1_infinite}      
    \\
   infinite-size scaling function for $\tilde \sigma (x)$ &$\tilde \sigma_{\infty} (x)$&
   \eqref{analyscalingfunction_ferro_2_infinite}         
    \\
   free energy density for the quantum ferromagnet & $f_{\Gamma}(m)$&
   \eqref{f_Gamma(m)Ferromagnet_Defition}
    \\
   infinite-size connecting point & $m_{\rm c}^{\infty}$&
   \eqref{GapFormula}
    \\
   mean energy density for the ground state at $\Gamma=\Gamma_{\rm c}^{\infty}$ & $e_{\rm c}^{\infty}$&
   \eqref{e_cfirstapperared}
 \end{longtable}
\end{center}

\end{appendices}

\section*{References}
\bibliography{finite}{}
\bibliographystyle{plain_url}

\end{document}